\documentclass[aps,prd,showpacs,preprint]{revtex4}
\usepackage{graphicx}
\newcommand{\nc}{\newcommand}
\nc{\beq}{\begin{equation}}   \nc{\eeq}{\end{equation}}
\nc{\bea}{\begin{eqnarray}}   \nc{\eea}{\end{eqnarray}}
\nc{\baa}{\begin{array}}      \nc{\eaa}{\end{array}}
\nc{\bit}{\begin{itemize}}    \nc{\eit}{\end{itemize}}
\nc{\ben}{\begin{enumerate}}  \nc{\een}{\end{enumerate}}
\nc{\bce}{\begin{center}}     \nc{\ece}{\end{center}}
\def\beqa{\begin{eqnarray}}
\def\eeqa{\end{eqnarray}}

\def\to{\rightarrow}

\def\lsim{\mathrel{\raise.3ex\hbox{$<$\kern-.75em\lower1ex\hbox{$\sim$}}}}
\def\gsim{\mathrel{\raise.3ex\hbox{$>$\kern-.75em\lower1ex\hbox{$\sim$}}}}
\def\ie{{\it i.e.\ }}

\def\non{\nonumber}
\begin{document}

\overfullrule 0pt
\preprint{\vbox{\vglue 0.6cm
\hbox{IFUSP-DFN/02-075}
\hbox{IFT-P.036/2002}
%\hbox{hep-ph/0206137}
}}
\vglue 0.2cm
\title{
 Constraining the absolute neutrino mass scale and Majorana 
 CP violating phases by future $0\nu\beta\beta$ decay 
 experiments               
} 
\author{H. Nunokawa$^1$}
\email{nunokawa@ift.unesp.br}
\author{W. J. C. Teves$^2$}
\email{teves@charme.if.usp.br}
\author{R. Zukanovich Funchal$^2$}
\email{zukanov@if.usp.br}
\affiliation{$^1$Instituto de F\'{\i}sica Te\'orica, 
Universidade Estadual Paulista, 
Rua Pamplona 145, 01405-900 S\~ao Paulo, Brazil \\
$^2$Instituto de F\'{\i}sica, Universidade de S\~ao Paulo 
C.\ P.\ 66.318, 05315-970 S\~ao Paulo, Brazil}

%\vspace{-0.4cm}
%%%%%%%%%%%%%%%%%%%%%%%%%%%%%%%%%%%%%%%%%%%%%%%%%%%%%%%%%%%%%%%%%%%
%\hfuzz=25pt
\begin{abstract}
 Assuming that neutrinos are Majorana particles, 
 in a three generation framework, 
 current and future neutrino oscillation experiments 
 can determine six out of the nine parameters 
 which fully describe the structure of 
 the neutrino mass matrix.
 We try to clarify the interplay among the remaining parameters, 
 the absolute neutrino  mass scale and two CP violating 
 Majorana phases, and how they can be accessed by future neutrinoless 
 double beta ($0\nu\beta\beta$) decay experiments, for the normal as 
 well as for the inverted order of the neutrino mass spectrum. 
 Assuming the oscillation parameters to be in the range presently 
 allowed by atmospheric, solar, reactor and accelerator neutrino 
 experiments, we quantitatively estimate the bounds on $m_0$, the 
 lightest neutrino mass, that can be inferred if the next 
 generation  $0\nu\beta\beta$ decay  experiments can probe the 
 effective Majorana mass ($m_{ee}$) down to $\sim$ 1 meV.
 In this context we conclude that in the case neutrinos are Majorana 
 particles: (a) if $m_0 \gsim 300$ meV, {\em i.e.}, within the 
 range directly attainable by future laboratory experiments 
as well as astrophysical observations, 
then $m_{ee} \gsim 30$ meV must be observed; 
 (b) if $m_0 < 300$ meV, results from future $0\nu\beta\beta$ decay 
 experiments combined with stringent bounds on the neutrino 
 oscillation parameters, specially the solar ones, will place much 
 stronger limits on the allowed values of $m_0$ than these direct 
 experiments. For instance, if a positive signal is observed around 
 $m_{ee}= 10$ meV, we estimate  $3 \lsim m_0/\text{meV} \lsim 65$  
 at 95 \% C.L.; on the other hand, if no signal is observed down 
 to $m_{ee}=10$ meV, then  $m_0\lsim 55$ meV  at 95 \% C.L.
\vspace{-0.4cm}
\end{abstract}
%\vskip2pc]

\maketitle
\newpage

%%%%%%%%%%%%%%%%%%%%%%%%%%%%%%%%%%%%%%%%%%%%%%%%%%%%%%%%%%%%%%%%%%%%%%
\section{Introduction}
\label{sec:intro}
%%%%%%%%%%%%%%%%%%%%%%%%%%%%%%%%%%%%%%%%%%%%%%%%%%%%%%%%%%%%%%%%%%%%%%
 During the last few years, a significant amount of information 
 on the size of the neutrino oscillation parameters have been gathered.
 Most of what we currently know about these parameters rely
 on evidences of neutrino flavor transformation that  
 have been collected by the experimental observations of solar~\cite{solarnu} 
 as well as of atmospheric neutrinos~\cite{atmnu}. 
 The evidences coming from solar neutrino data have been strengthened 
 by the recent neutral current measurement at Sudbury Neutrino 
 Observatory (SNO)~\cite{SNONC} whilst those coming 
 from atmospheric neutrinos have also been strongly supported 
 by the K2K  accelerator based neutrino oscillation experiment~\cite{k2k}. 
 Furthermore, the negative results of reactor experiments~\cite{reactors}
 also impose stringent limits on some oscillation parameters.

 Assuming that only three active neutrinos participate in oscillations
 in nature, independent of whether neutrinos are Dirac or Majorana 
 particles, current and future neutrino oscillation experiments can 
 determine at the most six out of the nine parameters which 
 completely describe the neutrino mass matrix, 
 {\em i.e.}, two mass squared 
 differences ($\Delta m_{12}^2$, $\Delta m_{23}^2$), three mixing angles 
 ($\theta_{12}$, $\theta_{13}$, $\theta_{23}$) and one CP violating 
 phase ($\delta$) which parametrize the Maki-Nakagawa-Sakata (MNS)~\cite{MNS} 
 leptonic mixing matrix.  
 See, for instance, Refs.~\cite{oscillation,kamland} for recent discussions 
 on the determination of these {\em oscillation} parameters. 

 However, if neutrinos are of the Majorana type, there remain three 
 {\em non-oscillation} parameters, which can not be accessed by 
 oscillation experiments. 
 They are the absolute neutrino mass scale, which can be taken as 
 the lightest neutrino mass, and two extra CP violating 
 Majorana phases~\cite{Schechter,Bilenky,Doi}. 
 It is well known that neutrinoless double beta ($0\nu \beta\beta$) decay 
 experiments can shed  light on these {\em non-oscillation} parameters. 

 The $0\nu \beta\beta$ decay is a process which can occur if and 
 only if neutrinos are Majorana particles~\cite{old}. 
 A positive signal of $0\nu \beta\beta$ decay always implies a 
 non-zero electron neutrino mass~\cite{shechter} 
 even if it is not induced by the exchange of a light neutrino 
 but by some other mechanism such as the one in supersymmetry 
 models with broken $R$-parity~\cite{susy}.  
 In this work, we assume the simplest possibility to be true, that 
 $0\nu \beta\beta$ decay process is induced only by 
 the exchange of a light neutrino. 

 It has been abundantly discussed in the literature 
 the relationship between the signals in $0\nu \beta\beta$ decay 
 experiments and oscillation phenomena, see for e.g., Ref.~\cite{0nuBB}.
 So far, a large amount of effort has been made to 
 constrain the oscillation parameters from the observation or  
 non-observation of $0\nu \beta\beta$ decay as well as to predict the 
 possible range of the effective Majorana mass in $0\nu \beta\beta$ decay 
 experiments,  $m_{ee}$, from the allowed range of oscillation 
 parameters~\cite{0nuBB}.

 In this paper, we take a different point of view.
 We examine how well we can constrain the three 
 {\em non-oscillation} parameters by future $0\nu \beta\beta$ 
 decay experiments, considering that the {\em oscillation} 
 parameters will be soon precisely determined 
 (or constrained) by current and future oscillation 
 experiments.
 We discuss the interplay among these parameters and the observable 
 signal in future  $0\nu \beta\beta$ decay experiments
 for the normal as well as for the inverted ordering of the neutrino 
 mass spectrum. In particular, presuming the oscillation parameters 
 to be in the range presently allowed by the atmospheric, the solar and the 
 reactor neutrino experiments, we examine what can be concluded 
 about these parameters in the case of either a positive or a 
 negative signal is obtained in future $0\nu \beta\beta$ decay 
 experiments~\cite{0nuBB-absolute}.

 So far, no convincing signal of $0\nu \beta\beta$ decay  
 has been observed, rather only an upper bound on $m_{ee}$
 \begin{equation}
 m_{ee} < 350 \ \text{meV},
 \label{eq:currentbound}
 \end{equation}
 which comes from the result of the Heidelberg-Moscow 
 Collaboration~\cite{upperbound} exists. 
 Recently, an experimental indication of the occurrence of  
 $0\nu \beta\beta$ decay has been announced~\cite{klapdor01} 
 but since this result seems to be controversial~\cite{comments},
 we do not discuss it in this work.

 There are many proposals for future $0\nu \beta\beta$ decay 
 experiments to go beyond the bound given in Eq.~(\ref{eq:currentbound}),
 those include GENIUS~\cite{GENIUS}, CUORE~\cite{CUORE}, 
 EXO~\cite{EXO}, MAJORANA~\cite{majorana} and NOON~\cite{NOON}. 
 It is expected that in the initial phase of the
 proposed GENIUS  experiment~\cite{GENIUS} the sensitivity to $m_{ee}$ 
 can be as low as $\sim$ 10 meV, going down to $\sim$ 2 meV, if 
 the 10 ton version of the experiment is implemented. 
 In this work, we will try to be optimistic and consider that future 
 experiments will eventually inspect  $m_{ee}$ down to $\sim$ 1 meV. 

 The absolute neutrino mass scale is also independently 
 constrained by the tritium decay experiments,
 which can directly measure the electron neutrino mass, obtaining  
 the upper bound~\cite{mainz}  
 \begin{equation}
 m_{\nu_e} < 2200 \ \text{meV}.
 \end{equation}
 The proposed KATRIN experiment aims to stretch the current sensitivity 
 down to $\sim 340$ meV~\cite{Katrin}. 
 We also take this into consideration in our discussions. 

 This paper is organized as follows. In Sec.~\ref{sec:formalism}, 
 we describe the theoretical framework on which we base our work. 
 We first discuss in Sec.~\ref{sec:m_0_dependence}, 
 the dependence of the $0\nu \beta\beta$ signal on the lightest 
 neutrino mass, and second in Sec.~\ref{sec:m_0_alpha_dependence} we 
 discuss how the dependence of $m_{ee}$ on $m_0$ is related to 
 the two CP phases, $\alpha_1$ and $\alpha_3$. 
 In Sec.~\ref{sec:m_0_s13_dependence} we discuss how 
 $m_{ee}^{\text{min}}$, the minimum possible value of  $m_{ee}$, 
 depends on $m_0$ and $\theta_{13}$. 
 Finally, in Sec.~\ref{sec:constraining_m_0}, 
 we discuss how the upper as well as the lower bounds on $m_0$ 
 depend on the solar neutrino oscillation parameters. 
 Sec.~\ref{sec:conclusions} is devoted to our discussions and 
 conclusions. 

%%%%%%%%%%%%%%%%%%%%%%%%%%%%%%%%%%%%%%%%%%%%%%%%%%%%%%%%%%%%%%%%%%%%%%
\section{The Formalism}
\label{sec:formalism}
%%%%%%%%%%%%%%%%%%%%%%%%%%%%%%%%%%%%%%%%%%%%%%%%%%%%%%%%%%%%%%%%%%%%%%

 In this section we discuss the theoretical frame work we will rely upon 
 in this work. 

\subsection{Mixing and mass scheme}
\label{subsec:mixingscheme}

 We consider mixing among three neutrino flavors as, 
\begin{equation}
\left[
\begin{array}{c}
\nu_e \nonumber\\
\nu_\mu \nonumber\\
\nu_\tau \nonumber\\
\end{array}
\right] 
= U 
\left[
\begin{array}{c}
\nu_1 \nonumber\\
\nu_2 \nonumber\\
\nu_3 \nonumber\\
\end{array}
\right], 
\end{equation}
 where $\nu_\alpha (\alpha = e,\mu,\tau)$ and 
 $\nu_i (i = 1,2,3)$ are the weak and the mass eigenstates, 
 respectively, and $U$ is the MNS~\cite{MNS} mixing matrix, 
 which can be parametrized as, 
\begin{equation}
\left[
\begin{array}{ccc}
c_{12}c_{13} & s_{12}c_{13} &   s_{13}e^{-i\delta}\nonumber\\
-s_{12}c_{23}-c_{12}s_{23}s_{13}e^{i\delta} &
c_{12}c_{23}-s_{12}s_{23}s_{13}e^{i\delta} & s_{23}c_{13}\nonumber\\
s_{12}s_{23}-c_{12}c_{23}s_{13}e^{i\delta} &
-c_{12}s_{23}-s_{12}c_{23}s_{13}e^{i\delta} & c_{23}c_{13}\nonumber\\
\end{array}
\right],
\label{MNSmatrix}
\end{equation}
 where $s_{ij}$ and $c_{ij}$, correspond to sine and cosine 
 of $\theta_{i,j}$.
 We define the neutrino mass-squared differences as
 $\Delta m^2_{ij} \equiv m^2_{j} - m^2_{i}$, where
 $\Delta m^2_{\odot} \equiv \Delta m^2_{12}$ is 
 relevant for the solutions to the solar neutrino problem, and 
 $\Delta m^2_{\text{atm}} \equiv |\Delta m^2_{23}| \simeq |\Delta m^2_{13}|$ 
 is relevant for the atmospheric neutrino observations. 

 Current atmospheric neutrino data~\cite{atmnu} indicate that 
 \begin{equation} 
 \Delta m^2_{\text{atm}} \equiv | \Delta m^2_{23} | 
 \simeq (2-4) \times 10^{-3} \ \text{eV}^2, \ \ 
 \sin^2 2\theta_{23} \simeq 0.9-1, 
 \label{ratm}
 \end{equation}
 which combined with nuclear reactor results~\cite{reactors} imply 
 \begin{equation} 
 \sin^2 \theta_{13} \lsim  0.02, 
 \label{rchooz}
 \end{equation}
 while the various solar neutrino experiment results
 strongly suggest that the so called large mixing angle (LMA) 
 MSW solution with parameter in the range~\cite{solarnu,SNONC,sol-recent}   
\begin{equation}
\Delta m^2_{\odot} \equiv \Delta m^2_{12} 
\simeq (2-40) \times 10^{-5} \ \text{eV}^2,  \ \ 
\tan^2 \theta_{12} \simeq 0.2-0.8, 
 \label{rsol}
\end{equation}
 will prevail as the explanation to the solar neutrino problem.
 We will admit throughout this paper that the actual values of these 
 parameters will be confirmed within the above ranges by 
 future neutrino oscillation experiments. Besides further constraining  
 the oscillation parameters given in Eqs.~(\ref{ratm})-(\ref{rsol}),  
 it is expected that these experiments also can probe the  
 CP violating phase $\delta$ and determine the neutrino 
 mass spectrum (sign of $\Delta m^2_{23}$)~\cite{oscillation}.

 In this work, we denote the lightest neutrino mass by $m_0$.  
 Then using $\Delta m^2_\odot$ and $\Delta m^2_{\text{atm}}$ as 
 defined above,  
 we can describe the two possible mass spectra as follows, 
\vglue 0.2cm
\noindent
 (a) Normal mass ordering: 
\vglue 0.1cm
\noindent
\begin{eqnarray}
 m_1  \equiv  & m_0, \non \\ 
 m_2  \equiv & \sqrt{m^2_0+\Delta m^2_\odot}, \non \\ 
 m_3  \equiv &\sqrt{m^2_0+\Delta m^2_\odot+\Delta m^2_{\text{atm}}}. 
\end{eqnarray}

\begin{figure}
\includegraphics[height=4.5cm]{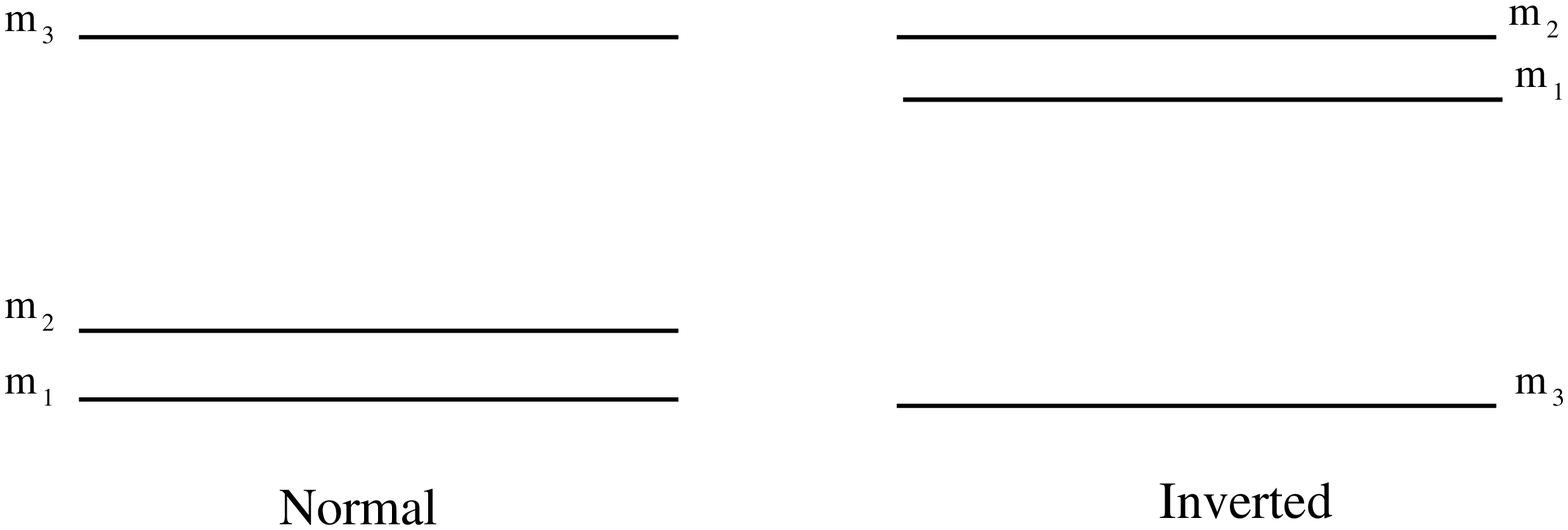}
\caption{
Mass ordering considered in this work. 
}
\label{fig:esquema}
\end{figure}

\vglue 0.1cm
\noindent
 (b) Inverted mass ordering: 
\vglue 0.2cm
\noindent
\begin{eqnarray}
 m_1  \equiv & \sqrt{m^2_0-\Delta m^2_\odot+\Delta m^2_{\text{atm}}}, \non \\ 
 m_2  \equiv & \sqrt{m^2_0+\Delta m^2_{\text{atm}}}, \non \\ 
 m_3  \equiv & m_0. 
\end{eqnarray}
 In this manner $\Delta m^2_{12} = \Delta m^2_\odot$ for both
 mass ordering and $\Delta m^2_{23} = \pm \Delta m^2_{\text{atm}}$
 where the $+(-)$ sign indicates normal (inverted) mass ordering.
 In Fig.~\ref{fig:esquema}, a schematic picture of the mass ordering 
 we consider here is shown. 

\subsection{Effective Majorana mass and $0\nu\beta\beta$}
\label{subsec:0nubb}

 Assuming that the $0\nu \beta\beta$ decay process occurs through 
 the exchange of a light $(m_\nu < 10$ MeV) neutrino,
 the theoretically expected half-life of the $0\nu \beta\beta$ decay, 
 $T^{0\nu}_{1/2}$, is given by~\cite{vogel}, 
\begin{equation}
[T^{0\nu}_{1/2}]^{-1}
= G^{0\nu} |M_{0\nu}|^2 m_{ee}^2,
\label{eq:halflife}
\end{equation}
 where $G^{0\nu}$ denotes the exact calculable phase space integral, 
 $M_{0\nu}$ consist of the sum of the Gamow-Teller and the Fermi 
 nuclear matrix elements defined as in Ref.~\cite{vogel} and $m_{ee}$ 
 is the effective Majorana mass defined in Eq.~(\ref{eq:m_ee}) below.

 It is known that the evaluation of the nuclear matrix elements
 suffers from a large uncertainty depending on the adopted method 
 used in the calculations. As we can see in Table II of Ref.~\cite{vogel}, 
 the evaluated half-lives for a given nucleus and a given value 
 of $m_{ee}$ typically vary within a factor of $\sim$ 10 comparing
 the largest and smallest predicted values. 
 This implies a factor of $\sim$ 3 difference between the minimum and 
 the maximum values of $m_{ee}$ when extracting it from the results of 
 $0\nu \beta\beta$ decay experiments, 
 which in fact directly measure or constrain not $m_{ee}$ but the value 
 of $T^{0\nu}_{1/2}$. This is clear from Eq. (\ref{eq:halflife}). 
 In Sec. \ref{sec:constraining_m_0}, we will consider for our
 estimations a somewhat optimistic uncertainty of a factor $\sim$ 2
 instead of 3, assuming future  improvements on the evaluation of the 
 nuclear matrix elements. 

 The effective Majorana mass, $m_{ee}$, is given by,
\begin{eqnarray}
m_{ee} 
& =  & | m_1 U^2_{e1} + m_2 U^2_{e2} + m_3 U^2_{e3}|, \nonumber \\
& =  & | m_1c^2_{12}c^2_{13}e^{2i\alpha_1}   + m_2s^2_{12}c^2_{13}
+ m_3s^2_{13}e^{2i\alpha_3} |, 
\label{eq:m_ee}
\end{eqnarray}
 where we have chosen to attach the CP violating phases to the first 
 and third elements.
%  as in Ref.~\cite{Frigerio:2002rd}. 
 Note that $\alpha_1$ and $\alpha_3$ must be understood as the relative 
 phases of $U_{e1}$ and $U_{e3}$ with respect to that of $U_{e2}$.
 The range of these phases are 
\begin{equation}
0 \le \alpha_1 \le \pi, \ \  0 \le \alpha_3 \le \pi. 
\label{eq:range1}
\end{equation}
 As it is known, the value of $m_{ee}$ can be perceived 
 as the norm of the sum of three vectors $\vec{m}_{ee}^{(1)}$, 
 $\vec{m}_{ee}^{(2)}$ and $\vec{m}_{ee}^{(3)}$ in the complex 
 plane whose absolute values are given by, 
\begin{eqnarray}
m_{ee}^{(1)}& \equiv & |\vec{m}_{ee}^{(1)}| \equiv 
|U^2_{e1}|m_1 = m_1c^2_{12}c^2_{13}, \nonumber \\
m_{ee}^{(2)}& \equiv & |\vec{m}_{ee}^{(2)}| \equiv 
|U^2_{e2}|m_2 = m_2s^2_{12}c^2_{13}, \nonumber \\
m_{ee}^{(3)}& \equiv & |\vec{m}_{ee}^{(3)}| \equiv 
 |U^2_{e3}|m_3 = m_3s^2_{13}.
\end{eqnarray}

 Explicitly, $m_{ee}$ is expressed as
\begin{eqnarray}
m_{ee}^2
 &= &
 [m_{ee}^{(1)}\cos 2\alpha_1+m_{ee}^{(2)}+m_{ee}^{(3)}\cos2\alpha_3]^2
 + [m_{ee}^{(1)}\sin2\alpha_1+m_{ee}^{(3)}\sin2\alpha_3]^2,
  \nonumber
  \\
&= &
[m_{ee}^{(1)}]^2+[m_{ee}^{(2)}]^2+[m_{ee}^{(3)}]^2 
 +2 \left\{ \right.
m_{ee}^{(1)}m_{ee}^{(2)} \cos 2\alpha_1  \nonumber \\
&& + m_{ee}^{(2)}m_{ee}^{(3)}\cos2\alpha_3 + 
m_{ee}^{(1)}m_{ee}^{(3)}\cos [2(\alpha_1-\alpha_3)]
\left. \right\}, \nonumber  \\
&= &
m_1^2c^4_{12}c^4_{13}+
m_2^2s^4_{12}c^4_{13}+
m_3^2s^4_{13}+ 
2\left\{ \right.
m_1m_2 c^2_{12}s^2_{12}c^2_{13}\cos 2\alpha_1 \nonumber  \\
&& +m_2m_3 s^2_{12}c^2_{13}s^2_{13}\cos2\alpha_3 
+ m_1m_3 c^2_{12}c^2_{13}s^2_{13}\cos [2(\alpha_1-\alpha_3)]
\left. \right\}.
\label{mee2}
\end{eqnarray}
 We can clearly see from Eq.~(\ref{mee2}) that $m_{ee}$ is invariant 
 under the transformation 
\begin{equation}
(\alpha_1,\alpha_3) \to (\pi - \alpha_1,\pi - \alpha_3), 
\end{equation}
 which allows us to further restrict the range of 
$(\alpha_1,\alpha_3)$, without loss of generality, to 
 \begin{equation}
0 \le \alpha_1 \le \pi, \ \  0 \le \alpha_3 \le \pi/2. \ \  
\label{eq:range2}
\end{equation}
 We note that $\alpha_1$ and/or $\alpha_3$ different 
 from 0 and $\pi/2$ imply CP violation. 

 The maximum possible value of $m_{ee}$, denoted by 
 $m_{ee}^{\text{max}}$, is given by, 
\begin{eqnarray}
m_{ee}^{\text{max}} 
&=& m_{ee}^{(1)}+m_{ee}^{(2)}+m_{ee}^{(3)}, \nonumber \\
&=&  (m_1c^2_{12}+m_2s^2_{12})c^2_{13}+m_3s^2_{13}.
\end{eqnarray}
 which occurs for $\alpha_1 = \alpha_3 = 0$. 
 On the other hand, the minimum possible value of $m_{ee}$ 
 is zero only when the three vectors  
 $\vec{m}_{ee}^{(i)}\ (i=1,2,3)$ can form a 
 triangle. This  can occur when the condition,
 \begin{equation}
 m_{ee}^{(i)} < \sum_{j\ne i} m_{ee}^{(j)} \  \ \ (i=1,2,3),
 \end{equation}
 is satisfied. 
 When these three vectors 
 can not form a closed triangle, which includes the case when one of 
 them is null, the minimum value is given by twice the length of the 
 largest vector minus the sum of the norm of all three vectors, 
\begin{equation}
 m_{ee}^{\text{min}} = 2 \max\{m_{ee}^{(1)},m_{ee}^{(2)},m_{ee}^{(3)}\}
 -[m_{ee}^{(1)}+m_{ee}^{(2)}+m_{ee}^{(3)}].
\end{equation}
 The values of ($\alpha_1$, $\alpha_3$) which lead to such a minimum are 
 ($\alpha_1$, $\alpha_3$) = ($\pi/2$, 0), ($\pi/2$, $\pi/2$) 
 or ($0$, $\pi/2$) when  respectively $m_{ee}^{(1)}$, $m_{ee}^{(2)}$ 
 or $m_{ee}^{(3)}$ is the largest contribution.

%%%%%%%%%%%%%%%%%%%%%%%%%%%%%%%%%%%%%%%%%%%%%%%%%%%%%%%%%%%%%%%%%%%%%%
\subsection{Some useful extreme limits}
\label{subsec:extreme}
%%%%%%%%%%%%%%%%%%%%%%%%%%%%%%%%%%%%%%%%%%%%%%%%%%%%%%%%%%%%%%%%%%%%%%
 To help the comprehension of our discussions in the following 
 sections, let us review here the approximated expressions for 
 $m_{ee}^{\text{max}}$ and  $m_{ee}^{\text{min}}$ for the two 
 extreme cases of the absolute neutrino mass scale: 
 vanishing $m_0$ and very large $m_0$ compared to 
 $\sqrt{\Delta m^2_{\text{atm}}}$.
 The neutrino oscillation 
 parameters are assumed to lie in the ranges  given in
 Eqs.~(\ref{ratm})-(\ref{rsol}).

 {\bf (i)} Vanishing $m_0$  limit:

 For the normal mass ordering we have, 
\begin{equation}
m_{ee}^{\text{min}}, m_{ee}^{\text{max}} 
 \simeq  m_{ee}^{(2)} \mp m_{ee}^{(3)} 
\simeq 
\sqrt{\Delta m^2_{\odot}} s^2_{12} c^2_{13} \mp
\sqrt{\Delta m^2_{\text{atm}}} s^2_{13}, 
\end{equation}
 where the $-(+)$ sign corresponds to 
 $m_{ee}^{\text{min}}$ $(m_{ee}^{\text{max}})$.  

 For the inverted mass ordering we have, 
\begin{eqnarray}
m_{ee}^{\text{min}}
& \simeq & m_{ee}^{(1)} - m_{ee}^{(2)} 
\simeq 
\sqrt{\Delta m^2_{\text{atm}}} \cos 2\theta_{12} c^2_{13}, 
\label{eq:smallm0_meemin_inverted}
\\
m_{ee}^{\text{max}}
& \simeq & m_{ee}^{(1)} + m_{ee}^{(2)} 
\simeq \sqrt{\Delta m^2_{\text{atm}}} c^2_{13}.
\label{eq:smallm0_meemax_inverted}
\end{eqnarray}

 {\bf (ii)} Large $m_0$ ($\gg \sqrt{\Delta m ^2_{\text{atm}}}$)  limit:

 For the normal as well as for the inverted mass ordering, 
\begin{eqnarray}
m_{ee}^{\text{min}}
& \simeq & m_{ee}^{(1)} -  m_{ee}^{(2)} -  m_{ee}^{(3)} 
 \simeq 
m_0 (c^2_{13} \cos2\theta_{12} - s^2_{13}), 
\label{eq:largem0meemin} \\ 
m_{ee}^{\text{max}}
& \simeq & m_{ee}^{(1)} +  m_{ee}^{(2)} +  m_{ee}^{(3)} 
 \simeq  m_0.
\label{eq:largem0meemax}
\end{eqnarray}

%%%%%%%%%%%%%%%%%%%%%%%%%%%%%%%%%%%%%%%%%%%%%%%%%%%%%%%%%%%%%%%%%%%%%%
\section{Dependence on the lightest neutrino mass}
\label{sec:m_0_dependence}
%%%%%%%%%%%%%%%%%%%%%%%%%%%%%%%%%%%%%%%%%%%%%%%%%%%%%%%%%%%%%%%%%%%%%%

\begin{figure}
\vglue -3.cm
\hglue -2.0cm
\includegraphics[height=21.cm,width=25.cm]{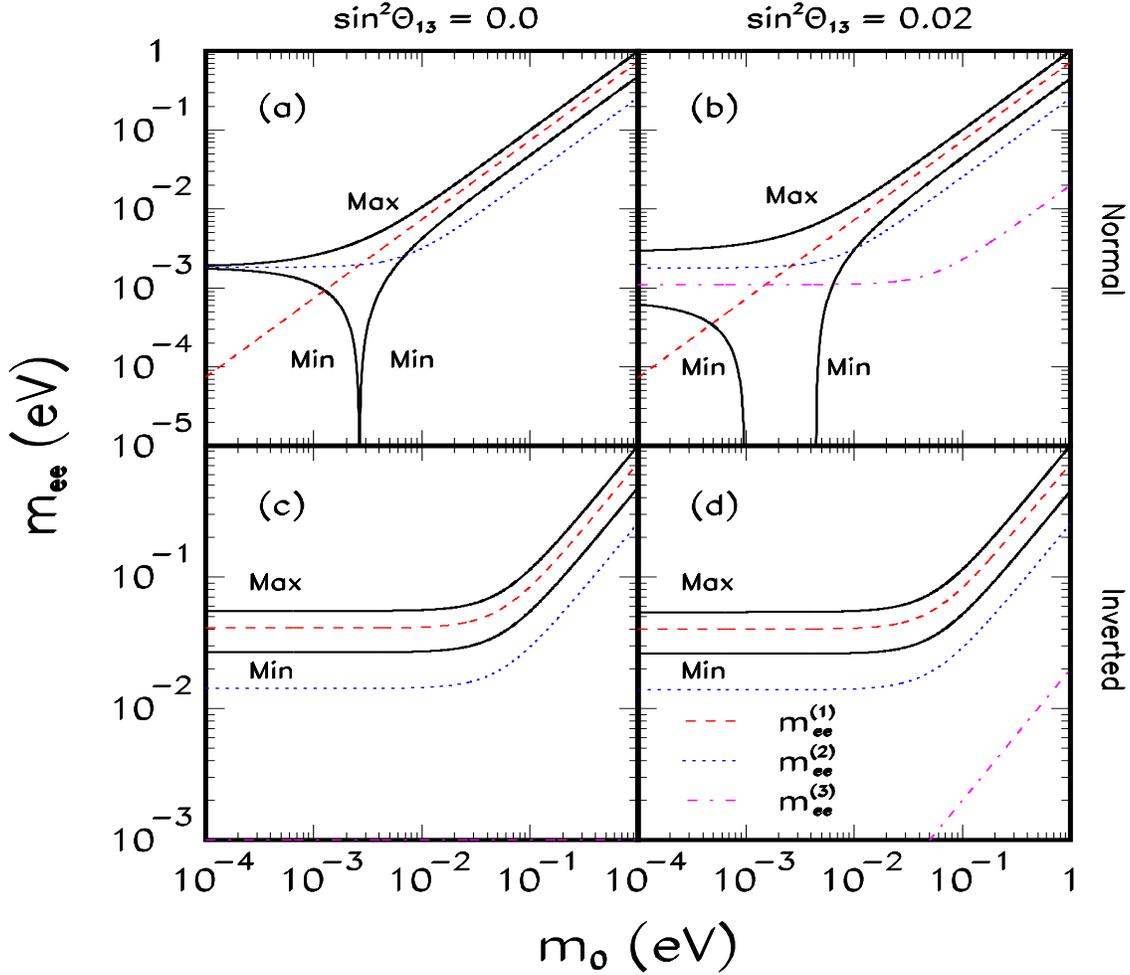}
\vglue -5.3cm
\caption{
Maximum and minimum possible values of $m_{ee}$ as a function 
of $m_0$ indicated by the thick solid curves for $\sin^2 \theta_{13}=0$
(left panels) and 0.02 (right panels) for 
normal (upper panels) as well as for inverted (lower panels) 
mass ordering. 
We have fixed the other mixing parameters as 
$\Delta m^2_{12} = 5 \times 10^{-5}$~eV$^2$, 
$\tan^2\theta_{12}$ = 0.35 and  
$|\Delta m^2_{23}| = 3 \times 10^{-3}$~eV$^2$.
 The individual contributions of $m_{ee}^{(1)}$, 
 $m_{ee}^{(2)}$ and $m_{ee}^{(3)}$ are also shown by 
 dashed, dotted and dash-dotted curves, respectively.
}
\label{fig:m_ee-m0}
\vglue -6.cm
\end{figure}

 In this section we examine how the effective Majorana mass, $m_{ee}$, 
 depends on the lightest neutrino mass, $m_0$, and 
 clarify  the importance of each of the individual 
 contributions, $m_{ee}^{(1)}$, $m_{ee}^{(2)}$ and $m_{ee}^{(3)}$
 to $m_{ee}$.  
 We present in Fig.~\ref{fig:m_ee-m0}
 for normal (upper panels) as well as inverted (lower panels) 
 mass ordering, the maximum and minimum values of $m_{ee}$
 as a function of $m_0$ for  vanishing $\theta_{13}$ (left panels) 
 and $\sin^2 \theta_{13}=0.02$ (right panels). 
 In the same plots we also show the individual 
 contributions of $m_{ee}^{(1)}$, $m_{ee}^{(2)}$ and $m_{ee}^{(3)}$ 
 by dashed, dotted and dash-dotted lines, respectively.

 Let us first discuss the case of normal mass ordering. 
 As we can see from the plots in the upper panels of Fig.~\ref{fig:m_ee-m0}, 
 for smaller values of $m_0$, $m_{ee}^{(2)}$ is the dominant contribution  
 whereas for larger values of $m_0$, $m_{ee}^{(1)}$ dominates over 
 the other contributions. 
  For the typical values of the oscillation parameters allowed by 
  the solar, atmospheric neutrino and reactor data, 
  $m_{ee}^{(3)}$ is almost always the smallest contribution to $m_{ee}$, 
  being subdominant in $m_{ee}$ and it can only be important when there 
  is a large 
  cancelation between $m_{ee}^{(1)}$ and $m_{ee}^{(2)}$. 

 It is clear that a strong cancelation in $m_{ee}$ 
 can occur if at least two of $m_{ee}^{(1)}$,  
 $m_{ee}^{(2)}$ and $m_{ee}^{(3)}$ are comparable in magnitude. 
 Just by comparing their magnitudes in the plots in  Fig.~\ref{fig:m_ee-m0}, 
 we can easily see for which values of $m_0$ a strong cancelation in 
 $m_{ee}$ can occur. 
 For the case if $\theta_{13}=0$, $m_{ee}$ can be zero only  
 at one particular value of $m_0$ 
 (see Fig.~\ref{fig:m_ee-m0} (a))
 whereas for the case if 
 $\theta_{13} \ne 0$ $m_{ee}$ can be zero for some range 
 of $m_0$ (see Fig.~\ref{fig:m_ee-m0}(b)).
 See Sec.~\ref{sec:m_0_s13_dependence} for more detailed 
 discussions about the dependence of $m_{ee}^{\text{min}}$ 
 on $m_0$ and $\theta_{13}$. 
 
 In the case of inverted mass ordering, the situation changes
 significantly. Here $m_{ee}^{(1)}>m_{ee}^{(2)}>>m_{ee}^{(3)}$ and 
 $m_{ee}^{\text{min}} \ne 0$  always  must satisfy the condition, 
\begin{equation}
 m_{ee}^{\text{min}} \gsim 
\sqrt{\Delta m^2_{\text{atm}}} \cos2\theta_{12} \sim 10 \ \ \text{meV},
\label{eq:meemin_inverted}
\end{equation}
 for any value of $m_0$ for current allowed parameters 
from solar and atmospheric neutrino data 
and no complete cancelation in $m_{ee}$ 
is expected as we can see clearly from the plots 
in the lower panels of Fig.~\ref{fig:m_ee-m0}. 
Therefore, if no positive signal of $0 \nu\beta\beta$ is
observed down to $\sim 10$ meV, inverted mass ordering 
can be excluded as long as neutrinos are Majorana particles. 

%%%%%%%%%%%%%%%%%%%%%%%%%%%%%%%%%%%%%%%%%%%%%%%%%%%%%%%%%%%%%%%%%%%%%%
\section{Dependence on the lightest neutrino mass and CP phases}
\label{sec:m_0_alpha_dependence}
%%%%%%%%%%%%%%%%%%%%%%%%%%%%%%%%%%%%%%%%%%%%%%%%%%%%%%%%%%%%%%%%%%%%%%

\begin{figure}
\vglue -2cm
\hglue -1.5cm
\includegraphics[height=27.cm]{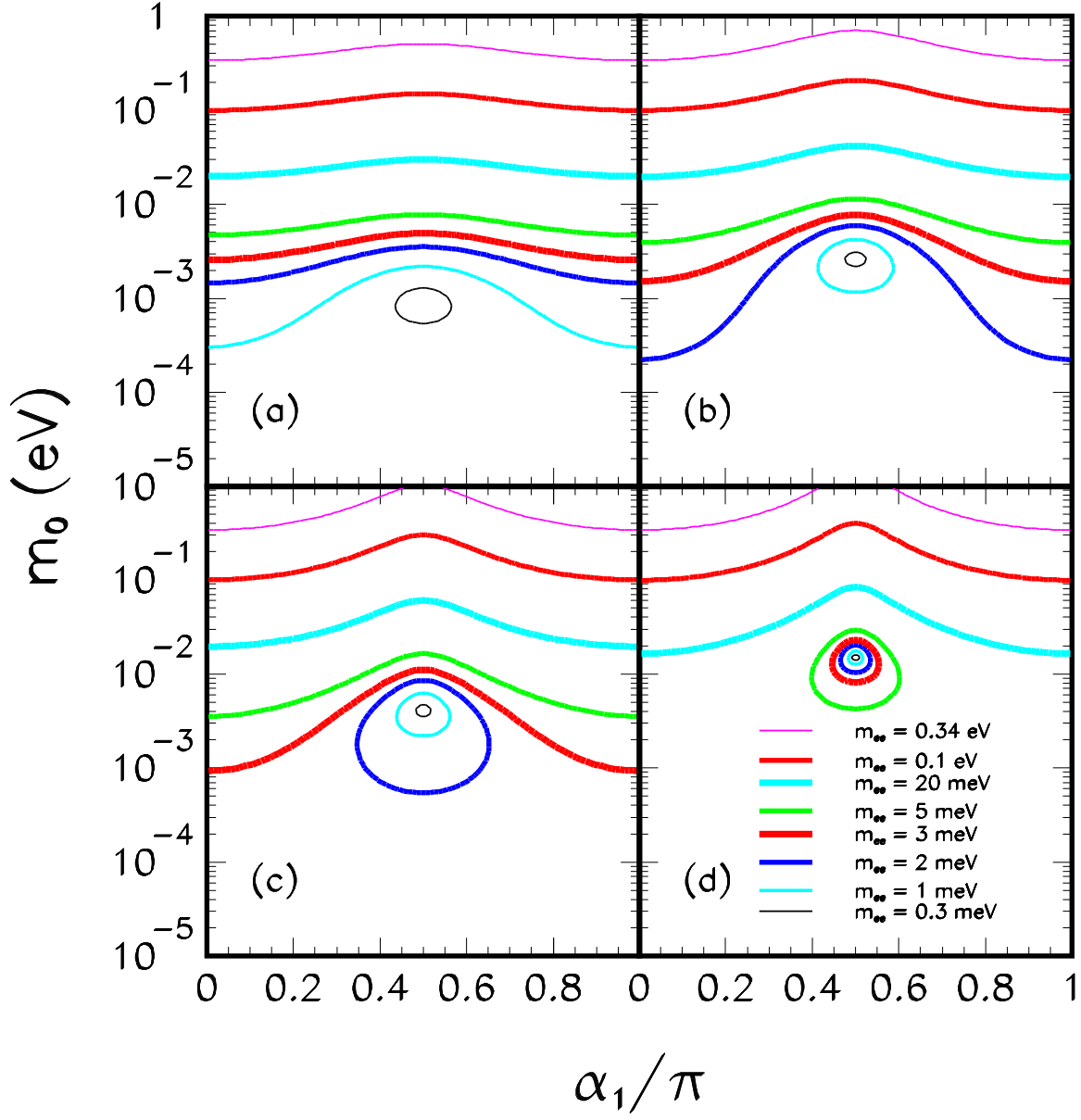}
\vglue -6.5cm
\caption{
Iso-contour plots of $m_{ee}$ for $\sin^2 \theta_{13}=0$ for the 
normal mass ordering in the $\alpha_1-m_0$ plane. 
We have fixed the other relevant mixing parameters as 
($\Delta m^2_{12}$, $\tan^2\theta_{12}$) = 
(2$\times 10^{-5}$~eV$^2$, 0.2)  in (a), 
(5$\times 10^{-5}$~eV$^2$, 0.35) in (b), 
(5$\times 10^{-5}$~eV$^2$, 0.5)  in (c) and 
(4$\times 10^{-4}$~eV$^2$, 0.6) in (d).
}
\label{fig:m_ee-alpha1-m_0_A}
\end{figure}

 Let us next discuss how the $m_{ee}$ dependence on $m_0$  is related to 
 the two CP phases, $\alpha_1$ and $\alpha_3$. 
 We present in Fig.~\ref{fig:m_ee-alpha1-m_0_A} iso-contour plots 
 of $m_{ee}$ in the  $m_0$-$\alpha_1$ plane for vanishing $\theta_{13}$, 
 for the normal mass ordering, for some values of the solar parameters 
 taken within the allowed ranges given in Eq.~(\ref{rsol}). 
 Note that, in this particular case,  
 there is no dependence on the  phase  $\alpha_3$ in our 
 choice of parametrization, which is clear from Eq.~(\ref{eq:m_ee}).

 Fist we note that in the plots, iso-contours are 
 symmetric with respect to $\alpha_1/\pi=0.5$.
 Second we note that if $m_{ee}$ is smaller
 than certain value, $m_{ee}^{\text{crit}}$, 
 the iso-contours are closed 
 which means that the possible values of both $m_0$ and  $\alpha_1$ 
 are bounded to some limited range, which does not include zero. 
 The critical value of $m_{ee}$ under which the contour is 
 a closed one is given by
\begin{equation}
m_{ee}^{\text{crit}} \simeq m_{ee}^{(2)} = \sqrt{\Delta m^2_\odot} s^2_{12} 
\simeq (1-10) \ \text{meV}, 
\end{equation}
 this dependence is illustrated in Fig.~\ref{fig:m_ee-alpha1-m_0_A}.
 If a positive $0 \nu \beta \beta$ signal is not observed down to 
 these values, this will imply either that $m_0$ as well as the CP 
 phase $\alpha_1$  are bounded to the limited range within the closed 
 contours shown in Fig.~\ref{fig:m_ee-alpha1-m_0_A} or that neutrinos 
 are of not Majorana type particles. 

\begin{figure}
\vglue -3cm
\hglue -2.5cm
\includegraphics[height=30.cm]{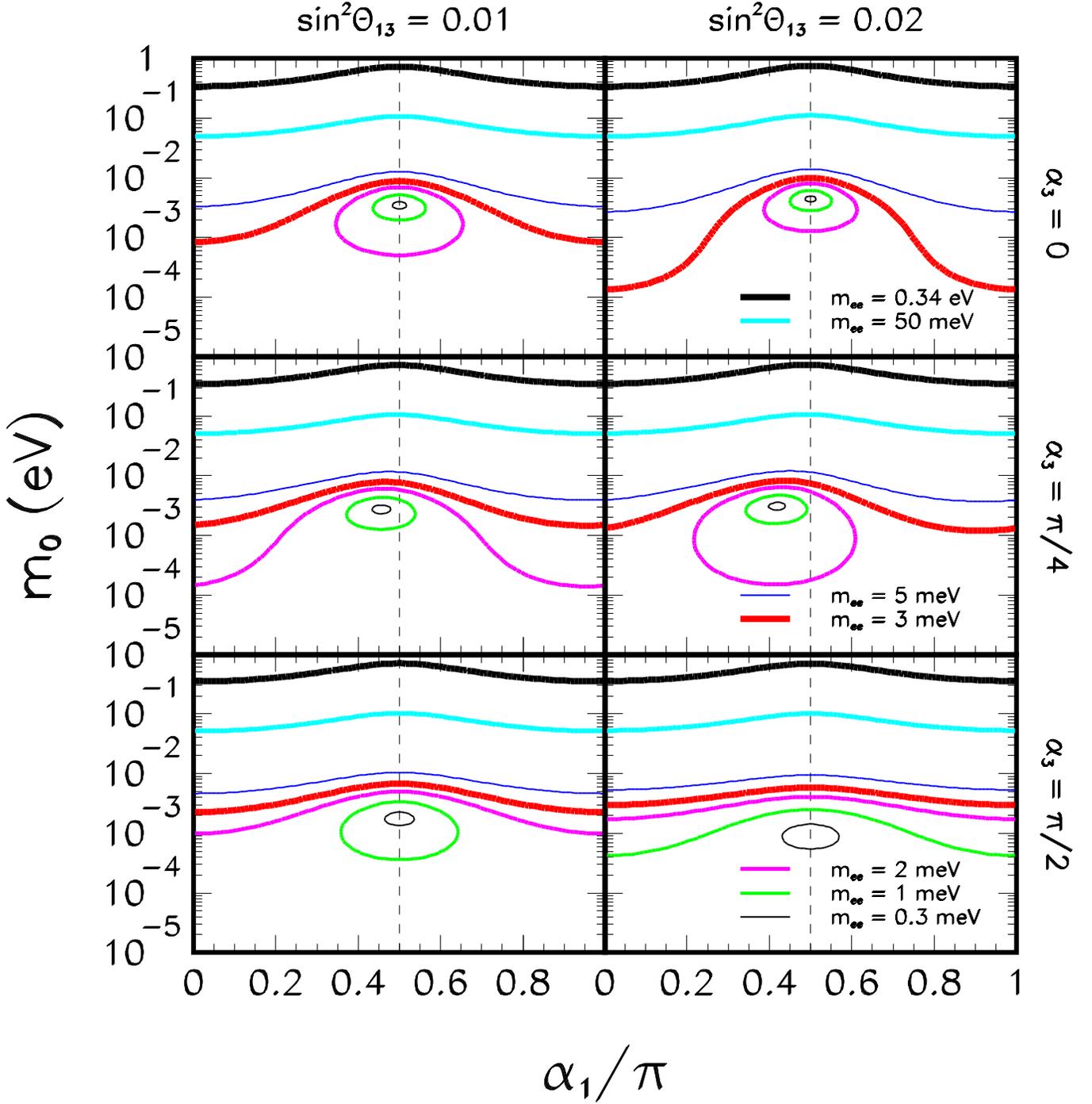}
\vglue -7.cm
\caption{
 Same as Fig.~\ref{fig:m_ee-alpha1-m_0_A}(b) but for 
 $\sin^2 \theta_{13}=0.01$ (left panels) and 0.02 (right panels),  
 for $\alpha_3 = 0$ (upper panels), $\pi/4$ (middle panels) 
 and $\pi/2$ (lower panels). The dashed vertical lines mark  
 $\alpha_1/\pi = 0.5$. 
}
\label{fig:m_ee-alpha1-m_0_B}
\end{figure}

 We show in Fig.~\ref{fig:m_ee-alpha1-m_0_B} the same 
 information as in  Fig.~\ref{fig:m_ee-alpha1-m_0_A}(b)
 but for $\sin^2\theta_{13}=0.01$ (left panels) 
 and 0.02 (right panels) for three different 
 values of $\alpha_3 = 0, \pi/4$ and $\pi/2$ 
 in the upper, middle and lower panels, respectively. 
 As we can see from these plots the qualitative behaviors 
 of the contours are very similar to those in 
 Fig.~\ref{fig:m_ee-alpha1-m_0_A}(b). 
 This is due to the fact that $m_{ee}^{(3)}$, which is the only term 
 that carries the  contributions of $\sin^2\theta_{13}$ and $\alpha_3$, 
 is subdominant compared to the other two elements in $m_{ee}$.  
 The effect of a non-zero $\alpha_3$ is to cause some displacement 
 of the position of the symmetry line of the plots from around 
 $\alpha_1/\pi=0.5$ to somewhat smaller values. 

 Let us here mention the case where $m_0$ can be independently 
 measured by another experiment such as the KATRIN~\cite{Katrin} 
 tritium decay one. In this case, it is possible to 
 constrain the CP phase $\alpha_1$ by comparing the measured values of
 $m_{ee}$ and $m_0$ provided that $m_0 \gsim 340$ meV, the
 maximum sensitivity of KATRIN. If a $0\nu\beta\beta$ decay experiment 
 measures $m_{ee}$  significantly smaller than $m_0$ measured by 
 KATRIN, this would imply a non-zero CP phase $\alpha_1$. 
 This is because for the $m_0$ values relevant for KATRIN, 
 $m_{ee} \sim m_0$ if $\alpha_1 \simeq 0$, 
 but $m_{ee}$ can be 
% very different from $m_0$, 
 as small as  $\sim 0.1 \times m_0$ if $\alpha_1 \sim \pi/2$ and 
 if the largest allowed value of $\theta_{12}$ from 
 the current LMA allowed region is realized 
(see Eqs.~(\ref{eq:largem0meemin}) and (\ref{eq:largem0meemax})).
 However, it would still be difficult to say something definite about 
 the value of $\alpha_1$ for 10 $\lsim m_0/\text{meV} \lsim$ 340.

 Our plots in Figs.~\ref{fig:m_ee-alpha1-m_0_A} and 
 \ref{fig:m_ee-alpha1-m_0_B} are in agreement with the conclusion 
 presented in Ref.~\cite{barger}, that is, either a positive or a negative 
 results in a $0\nu \beta\beta$ decay experiment can constrain $m_0$ and 
 $\alpha_1$ but, since the possible values of $\alpha_1$  will always 
 include $\pi/2$, a non-zero value of $\alpha_1$  cannot be interpreted 
 as an evidence of CP violation, even if a positive signal of 
 $0\nu \beta\beta$ decay is observed. 
 Unfortunately, to be able to say anything 
 more definite on the CP phase, independent precise information on $m_0$ is 
 unavoidable.
 Moreover, nothing can be concluded about the value of $\alpha_3$.

\begin{figure}
\vglue -2.cm
\hglue -1.5cm
\includegraphics[height=27.cm]{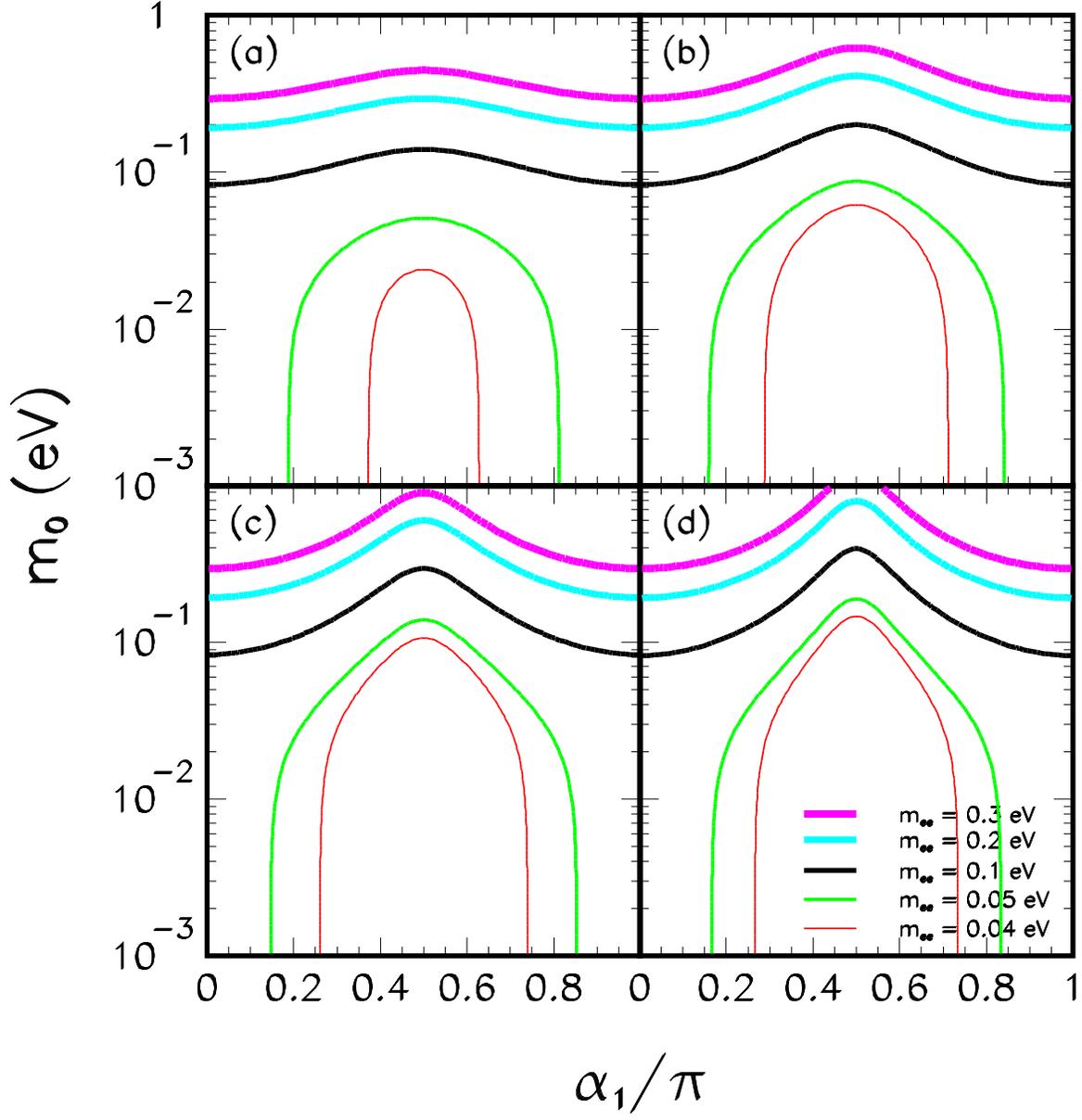}
\vglue -6.cm
\caption{
 Same as Fig.~\ref{fig:m_ee-alpha1-m_0_A} but for the inverted mass ordering. 
 Here we have fixed the atmospheric mass scale to  
 $\Delta m^2_{23} = -5 \times 10^{-3}$~eV$^2$ in (a), 
 to $\Delta m^2_{23} = -3 \times 10^{-3}$~eV$^2$ in (b) and (c) 
 and to $\Delta m^2_{23} = -1.3 \times 10^{-3}$~eV$^2$ in (d). 
}
\label{fig:m_ee-alpha1-m_0_C}
\end{figure}

 We show in Fig.~\ref{fig:m_ee-alpha1-m_0_C} the same information 
 as in Fig.~\ref{fig:m_ee-alpha1-m_0_A} for the inverted mass ordering. 
 Since there is no significant 
 dependence on $\theta_{13}$ or on $\alpha_3$ in this case, 
 we show only the curves for vanishing $\theta_{13}$.
 We note from these plots that there is no lower bound 
 for $m_0$ as long as $m_{ee} \lsim 50$ meV, and moreover, 
 $\alpha_1$ is less constrained if compared to 
 the case of normal ordering, independently 
 of the values of the other neutrino oscillation parameters.

 For the case of the normal mass ordering, we have also 
 investigated if the uncertainties in the determination of the
 solar parameters as well as on $m_{ee}$ can wash out the
 determination  of a non-zero CP phase $\alpha_1$, expected by the
 closed contours in Figs.~\ref{fig:m_ee-alpha1-m_0_A} and 
 \ref{fig:m_ee-alpha1-m_0_B}.
 For four different central values of $m_{ee}$, assuming 30\%
 uncertainty in their determination (see Sec.~\ref{sec:constraining_m_0} 
 for a detailed explanation), we have obtained the region in the 
($\Delta m^2_{12}$, $\tan^2 \theta_{12}$) plane where $\alpha_1$ can
 be constrained to a non-zero value for 
 (i) $\sin^2 \theta_{13} = 0$ and (ii) $\sin^2 \theta_{13} = 0.02$.
 This is shown in Fig.~\ref{fig:nova}, where we also have assumed 
 for each point in the ($\Delta m^2_{12}$, $\tan^2 \theta_{12}$) 
 plane a 10 \% uncertainty in the determination of these two parameters.
 In this plot we have indicated by crosses the set of solar parameters used 
 in Fig.~\ref{fig:m_ee-alpha1-m_0_A}(a)-(d). 

 We observe that for the case of $\theta_{13} = 0$ 
 (see Fig.~\ref{fig:nova} (i)),  
 determination of a non-zero CP phase $\alpha_1$ is possible 
 as long as one can reach the sensitivity $m_{ee} \lsim 5$ meV  
 for the paramerter set (d) and $m_{ee} \lsim 1 $ meV for 
 sets (b) and (c).
 On the other hand, in the case where $\sin^2 \theta_{13} = 0.02$, 
 as we can see from Fig.~\ref{fig:nova} (ii), 
 a better sensitivity in $m_{ee}$ is required to establish 
 non-zero $\alpha_1$ value for the same parameter set. 
 This is because when 
 $\theta_{13}$ is non-zero, third element $m_{ee}^{(3)}$, 
 which contains another CP phase, $\alpha_3$, whose value
 is assumed to be unknown, comes into play and this could 
 wash out more efficiently the determination of non-zero 
 $\alpha_1$ when compared to the case where $\theta_{13} = 0$. 

%%%%%%%%%%%%%%%%%%%%%%%%%%%%%%%%%%%%%%%%%%%%%%%%%%%%%%%%%%%%%%%%%%%%%%
\section{Dependence on the lightest neutrino mass and $\theta_{13}$}
\label{sec:m_0_s13_dependence}
%%%%%%%%%%%%%%%%%%%%%%%%%%%%%%%%%%%%%%%%%%%%%%%%%%%%%%%%%%%%%%%%%%%%%%

\begin{figure}
\vglue -3.cm
\hglue -3cm 
\includegraphics[height=30.cm]{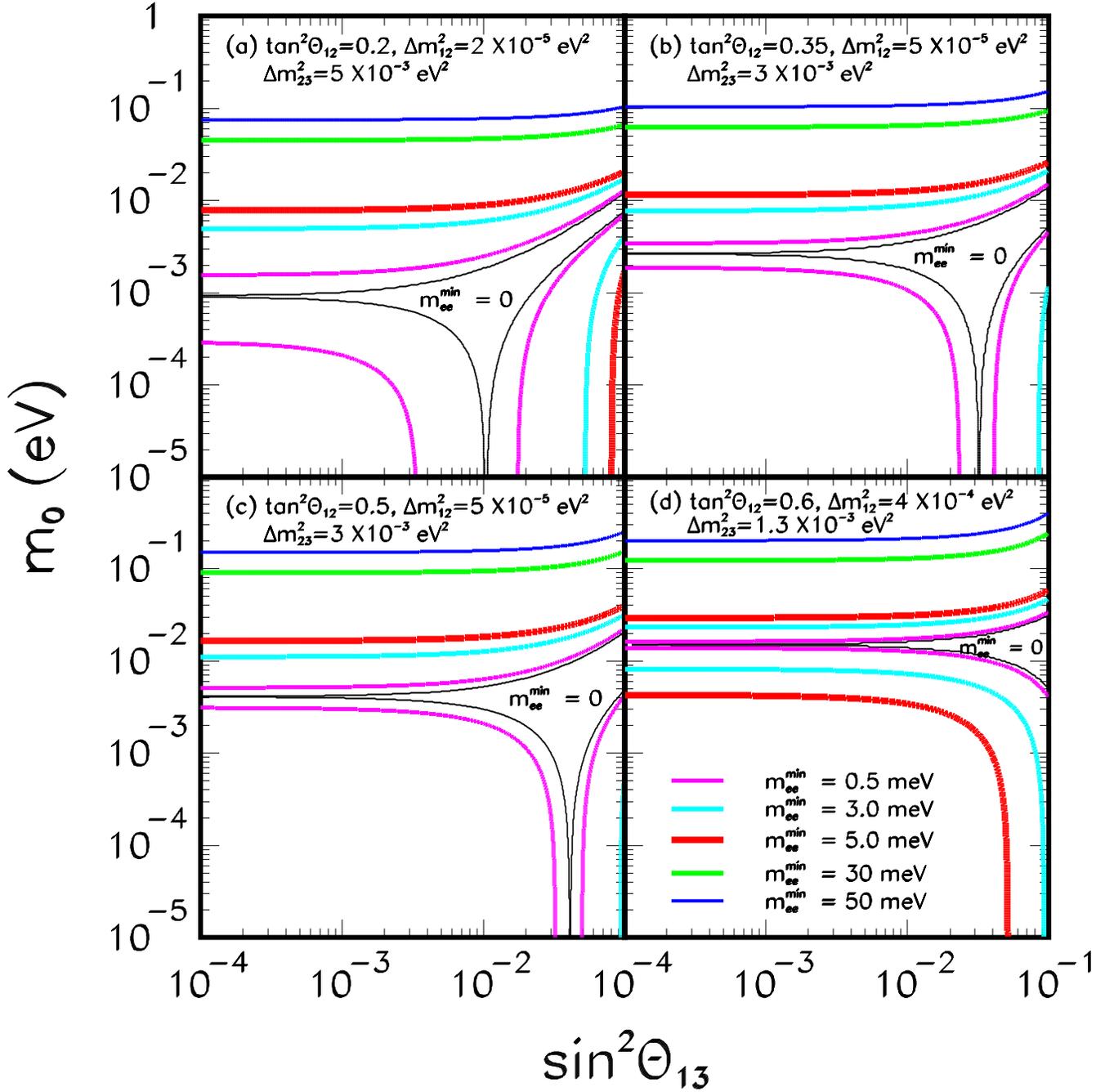}
\vglue -7.cm
\caption{
Iso-contour plots of $m_{ee}^{\text{min}}$ in 
the $s_{13}^2-m_0$ plane for different 
choices of the mixing parameters for the normal
mass ordering. We note that in the region delimited 
by thin solid curves $m_{ee}^{\text{min}} = 0$. 
}
\label{fig:m_ee-s13-m_0}
\end{figure}

 In this and the next sections we focus on the relation among  $m_{ee}$ 
 and some of the yet undetermined mixing parameters. 
 Let us start by discussing  the dependence of  $m_{ee}^{\text{min}}$ 
 on $m_0$ and $\theta_{13}$. 
 Here we discuss only the case of normal ordering since the dependence
 on $\theta_{13}$ for the inverted case is quite small. 
 We present in Fig.~\ref{fig:m_ee-s13-m_0}, 
 the iso-contour plots of $m_{ee}^{\text{min}}$ in the $s^2_{13}$-$m_0$ 
 plane for some  different choices of the oscillation parameters. 

  For vanishing $\theta_{13}$, one can have  $m_{ee}^{\text{min}} = 0 $ 
  if $m_{ee}^{(1)} = m_{ee}^{(2)}$, \ie, $m_1c^2_{12} = m_2s^2_{12}$, 
  which is equivalent to, 
 \begin{equation}
 m_0 = \frac{ s^2_{12} }{ \sqrt{|\cos2\theta_{12}|}}
 \sqrt{\Delta m^2_\odot} 
 \sim 3 \ \text{meV},
 \label{eq:critical-m_0}
 \end{equation}
 where we have computed the numerical estimate using the best 
 fitted values of the parameters from the latest solar neutrino data
 (see Fig.~\ref{fig:m_ee-m0}(a) and Fig.~\ref{fig:m_ee-s13-m_0}(b)).
 Within the current allowed range given in Eq.~(\ref{rsol}), 
 the possible values of $m_0$ for which we can expect
 strong cancelation are in the range 
 $m_0 \simeq (0.9-12) \ \text{meV}$, as we can confirm with the plots 
 in Fig.~\ref{fig:m_ee-s13-m_0}.  

 For non-zero values of $\theta_{13}$, as we can see clearly from 
 Fig.~\ref{fig:m_ee-s13-m_0}, 
 $m_{ee}^{\text{min}}$ can take zero for some range of $m_0$. 
 We can also see that there is a critical value of $\theta_{13}$ for 
 which 
 $m_{ee}^{\text{min}}$ is zero with vanishing $m_0$. 
 Such a value of $\theta_{13}$ can be easily estimated 
 by solving $m_2c^2_{13}s^2_{12} = m_3s^2_{13}$ with $m_0\rightarrow 0$ 
 which is  equivalent to, 
 \begin{equation}
 s^2_{13} \simeq  
 \sqrt{ 
 \frac{\Delta m^2_\odot}{\Delta m^2_{\text{atm}}}
 }
 s^2_{12} \sim 0.03,  
 \label{eq:s13crit}
 \end{equation}
 for the best fitted parameters from solar as well as atmospheric 
  neutrino data. 
 We note that this is close to the current upper 
 bound on $s^2_{13}$ allowed by the CHOOZ result~\cite{reactors}.  
 Letting $\Delta m^2_\odot$, 
 $\Delta m^2_{\text{atm}}$  and $s^2_{12}$ take any value in the 
 region consistent with the solar and atmospheric neutrino observations 
 given in Eq.~(\ref{ratm}) and (\ref{rsol}), the range  of
 $s^2_{13}$ for which strong cancelation can occur
 is $s^2_{13} \simeq  0.01-0.20$, which is again consistent 
 with our results in Fig.~\ref{fig:m_ee-s13-m_0}. 

 We observe that if $s^2_{13}$ is smaller than the critical 
 value given in Eq.~(\ref{eq:s13crit}), the value of $m_0$ can be 
 strongly constrained to some limited range around the value of $m_0$ 
 given in Eq.~(\ref{eq:critical-m_0}), provided that future 
 $0\nu \beta \beta$ experiment can probe a $m_{ee}$ value as small as 
 $\sim \sqrt{\Delta m^2_{\odot}}s^2_{12}$ independent of whether a positive 
 or negative signal of $0\nu \beta \beta$ is observed. 

%%%%%%%%%%%%%%%%%%%%%%%%%%%%%%%%%%%%%%%%%%%%%%%%%%%%%%%%%%%%%%%%%%%%%%
\section{Constraining $m_0$ using solar neutrino data}
\label{sec:constraining_m_0}
%%%%%%%%%%%%%%%%%%%%%%%%%%%%%%%%%%%%%%%%%%%%%%%%%%%%%%%%%%%%%%%%%%%%%%
  Finally, let us discuss how we can constrain $m_0$ using the solar 
  neutrino parameters. Let us first analyze the case where a positive 
  signal of $0\nu \beta \beta$ is observed. 
  The value of $m_{ee}$ has to be extracted from the experimental 
  measured half-life, $T^{0\nu}_{1/2}$, of the decaying parent nucleus 
  by comparison with the theoretical prediction which rely on nuclear 
  matrix element calculations. This means that the experimental value of 
  $m_{ee}$ has to be expressed as an interval obtained using the
  maximum and minimum matrix element predictions.
  
  As mentioned in Sec.~\ref{subsec:0nubb}, 
  typically, there is a factor 3 difference among the 
  results of the matrix element evaluations according to different model 
  assumptions~\cite{vogel}. 
  In order to take such large theoretical uncertainty into account 
  in our estimations, we will first assume that the experimental
  measured value of $m_{ee}$ will be extracted using the mean between 
  the minimum and maximum values of the matrix element calculations,
  then attach 30 \% uncertainty around this value, which correspond to
  a factor $\sim$ 2 between the smallest and the largest theoretically
  allowed $m_{ee}$ values for a given value of the observed half-life.
  This is a somewhat optimistic but reasonable assumption.
  Hopefully improvements on the understanding of the underlining 
  nuclear physics effects can decrease this uncertainty even further.

  In Fig.~\ref{fig:m_0sol_norm_posi0} we show the iso-contours 
  of upper ($m_0^{\text{max}}$) and lower 
  ($m_0^{\text{min}}$) bounds on $m_0$ in units of meV in 
  the $\tan^2\theta_{12}$-$\Delta m^2_{12}$ plane for the case where 
  a positive signal of $0\nu\beta\beta$ is observed with central 
  values $m_{ee}$ = 10, 5, 3  and 1 meV with a 30 \% uncertainty. 
  In these plots we have used   $\Delta m^2_{23} = 3 \times 10^{-3}$ eV$^2$ 
  and  $\sin^2 \theta_{13}=0$. 
  We do not present plots for $m_{ee}> 10$ meV, since in these cases 
  the upper and lower bound can be analytically estimated as we will 
  see below. 
  
  When $m_0 > \sqrt{\Delta m^2_{\text{atm}} }$ the upper bound on
  $m_0$ does not depend much on $\Delta m^2_{\odot}$ but essentially 
  only on $\theta_{12}$ (see Eq.~(\ref{eq:largem0meemin}) 
  in Sec.~\ref{subsec:extreme}), it is given as, 
  \begin{equation}
  m_0^{\text{max}} \sim 
  \frac{m_{ee}}{\cos2\theta_{12}}, 
  \label{eq:m_0_max} 
  \end{equation}
  and the lower bound $m_0^{\text{min}} \sim m_{ee}$, independent 
  of the solar parameters in the region compatible with the LMA MSW 
  solution  to the solar neutrino problem. 

  We observe that, as can be seen in Fig.~\ref{fig:m_0sol_norm_posi0},
  that as $m_{ee}$ decreases, the upper bound lines, which 
  mainly depend on $\tan^2 \theta_{12}$, are shifted to larger values 
  of $\theta_{12}$, decreasing $m_0^{\text{max}}$ for a given set of 
  $(\tan^2\theta_{12}, \Delta m^2_{12})$. 
  The lower bound lines, on the other hand, depend more on 
  $\Delta m^2_{12}$ and there are some regions where no lower bound 
  is obtained. 
  For $m_{ee} \gsim 10 $ meV or $m_{ee} \lsim 1$ meV there is 
  always a lower bound found inside the currently allowed 
  LMA MSW region, whereas 
  for $ 1 \lsim m_{ee}/\text{meV} \lsim 10$ this is not true. 
  The appearance of these no lower bound bands comes from the fact 
  the in these regions the solar mass scale alone is sufficient to 
  explain the positive signal observed, even for vanishing $m_0$. 
  
  In Fig.~\ref{fig:m_0sol_norm_posi2} we repeat the same exercise
  but  for $\sin^2 \theta_{13}=0.02$. The most significant effect 
  of a non zero $\sin^2 \theta_{13}$ is the increase of the size 
  of the no lower bound band, which can in some cases stretch over 
  the entire LMA allowed 
  region, otherwise the behavior of the upper and lower bound lines 
  is qualitatively similar to the previous case.   

  Let us next consider the case where no positive signal is observed. 
  In Fig.~\ref{fig:m_0sol_norm_nega0} we present the same 
  information as in Fig.~\ref{fig:m_0sol_norm_posi0} but for 
  the case where no positive signal of $0\nu\beta\beta$ 
  is observed down to $m_{ee}$ = 10 meV (a), 5 meV (b), 3 meV (c) 
  and 1 meV (d). 
  It is assumed that when no positive signal is observed, 
  for a given bound on the half-life time, 
  the bounds on $m_{ee}$ are extracted using the smallest 
  nuclear matrix element prediction which lead to 
  the largest $m_{ee}$ value (see Eq.~(\ref{eq:halflife})). 
  We can see that qualitative behavior of the iso-contours
  for the upper bound is similar to that in 
  Fig.~\ref{fig:m_0sol_norm_posi0} but it is different for 
  the trend of the lower bound curves. 
  Compared to the case where positive $0\nu\beta \beta$ signal
  is obtained, it is more difficult to put lower bound on $m_0$. 
  This can be easily understood from Fig.~\ref{fig:m_ee-m0} (a). 
  We note that unless we can constrain $m_{ee}$ down to 
  $\sim 3$ meV, no lower bound on $m_0$ is obtained. 
  In Fig.~\ref{fig:m_0sol_norm_nega2} we present 
  the same information as in Fig.~\ref{fig:m_0sol_norm_nega0} 
  but for $\sin^2 \theta_{13}=0.02$. 
  Again, the qualitative behavior of the iso-contours is similar
  to the case of $\sin^2 \theta_{13}=0$. 
  The difference from the previous case presented in 
  Fig.~\ref{fig:m_0sol_norm_nega0} is that 
  the constraint on $m_0$ become somewhat weaker, 
  leading to larger upper bounds and smaller lower bounds 
  for a given set of $(\tan^2\theta_{12}, \Delta m^2_{12})$. 

  Finally, let us also comment the case of inverted ordering. 
  For this case, we can easily estimate the upper as well as
  lower bound from the analytic expressions as well as from 
  the lower panels of Fig.~\ref{fig:m_ee-m0}. 
  Let us look at Fig.~\ref{fig:m_ee-m0} (c) and (d). 
  First of all, if the inverted ordering is the case, the observed 
  value of $m_{ee}$ in $0\nu\beta\beta$ decay experiment must be
  larger than  the value given in Eq.~(\ref{eq:meemin_inverted}), 
  as already mentioned in Sec.~\ref{sec:m_0_dependence}. 
  If the observed $m_{ee}$ is smaller than 
  $\sqrt{\Delta m^2_{\text{atm}}} c^2_{13}$ 
  (see Eq.~(\ref{eq:smallm0_meemax_inverted}))
  then there is no lower bound on $m_0$ whereas
  the upper bound is given by the same expression 
  for the case of normal ordering, by Eq.~(\ref{eq:m_0_max}).   
  If the observed $m_{ee}$ value is larger than 
  $\sqrt{\Delta m^2_{\text{atm}}} c^2_{13}$, then 
  the upper as well as lower bounds are given 
  by the same expression as in the case of normal ordering,
  which is already discussed in this section. 

%%%%%%%%%%%%%%%%%%%%%%%%%%%%%%%%%%%%%%%%%%%%%%%%%%%%%%%%%%%%%%%%%%%%%%
\section{Discussions and conclusions}
\label{sec:conclusions}
%%%%%%%%%%%%%%%%%%%%%%%%%%%%%%%%%%%%%%%%%%%%%%%%%%%%%%%%%%%%%%%%%%%%%%
 We have studied how the {\em non-oscillation} neutrino 
 parameters which can not be extracted from the oscillation analysis, 
 the absolute neutrino mass scale and two CP violating Majorana phases, 
 can be accessed by the positive or negative signal 
 of future $0\nu \beta \beta$ decay experiments. 
 We have carried out this by choosing various different set 
 of mixing parameters which are varied within the parameter region 
 currently allowed by the solar, atmospheric as well as by reactor neutrino 
 experiments. 

 In the future, the KATRIN experiment expects to directly inspect $m_0$ 
 down to $\sim$ 340 meV~\cite{Katrin}, while there are several proposed 
 astrophysical measurements on the temperature perturbation
 in the early universe imprinted in the cosmic microwave background radiation, 
 such as the ones that can be performed by MAP (Microwave Anisotropy 
 Probe)~\cite{MAP} and Planck~\cite{Planck}, which can probe $m_0$ down 
 to $\sim$ 300 meV~\cite{cosmology}, where the expected sensitivity suffers 
 from the uncertainty coming from cosmological parameters. 
 It is expected that future supernova neutrino measurements 
 can probe $m_0$ down to at most $\sim$ (2-3) eV~\cite{supernova}.

 Our analysis permit us to conclude that if these experiments 
 measure $m_0 \gsim 300$ meV, then either  a positive signal 
 of $0\nu\beta\beta$ decay compatible with $m_{ee} \gsim 30$ meV 
 must be observed in the near future or neutrinos are Dirac particles. 
 This is simply because $m_{ee}^{\text{min}}
 \sim m_0\cos 2 \theta_{12}$ (see Eq. (\ref{eq:largem0meemin}) 
 in Sec.~\ref{subsec:extreme}) and 
 $\cos 2 \theta_{12}\gsim 0.1$ from the current 
 allowed LMA parameter region given in Eq. (\ref{rsol}). 

 On the other hand, if these experiments do not observe any positive 
 signal of  $m_0 \gsim 300$ meV, results from future $0\nu\beta\beta$ decay 
 experiments combined with more precise values of the neutrino 
 oscillation parameters, specially the solar ones ($\Delta m^2_{12}$, 
  $\tan^2 \theta_{12}$), 
 which can be precisely determined by the KamLAND experiment~\cite{kamland}, 
 will place more stringent bounds on 
 $m_0$ for the Majorana case provided that the sensitivity 
 of $m_{ee} \lsim$ 30 meV is achieved. 
 To be more specific, if a positive signal is observed around 
 $m_{ee}=10$ meV (assuming a 30\% uncertainty 
 on the determination of $m_{ee}$), we estimate  
 $3 \lsim m_0/\text{meV} \lsim 65$  at 95 \% C.L.; on the other hand, 
 if no signal is observed down to $m_{ee}=10$ meV, 
 then  $m_0\lsim 55$ meV  at 95 \% C.L.
 Allowing for a more optimistic sensitivity, a positive signal  
 observed around $m_{ee}=3$ meV or no signal seen down to 3 meV
 would mean  $m_0 \lsim 25$ meV at 95 \% C.L. 
 These bounds can be improved by a better determination 
 of $\tan^2 \theta_{12}$, $\Delta m^2_{12}$ and  
 $\sin^2 \theta_{13}$ as can be clearly seen in 
 Figs.~\ref{fig:m_0sol_norm_posi0}-~\ref{fig:m_0sol_norm_nega2}, 
 as well as by the reduction of the uncertainties in the theoretical 
 calculations of the nuclear matrix elements. 

\begin{figure}
\vglue -1.cm
\hglue -1.0cm
\includegraphics[height=18.cm]{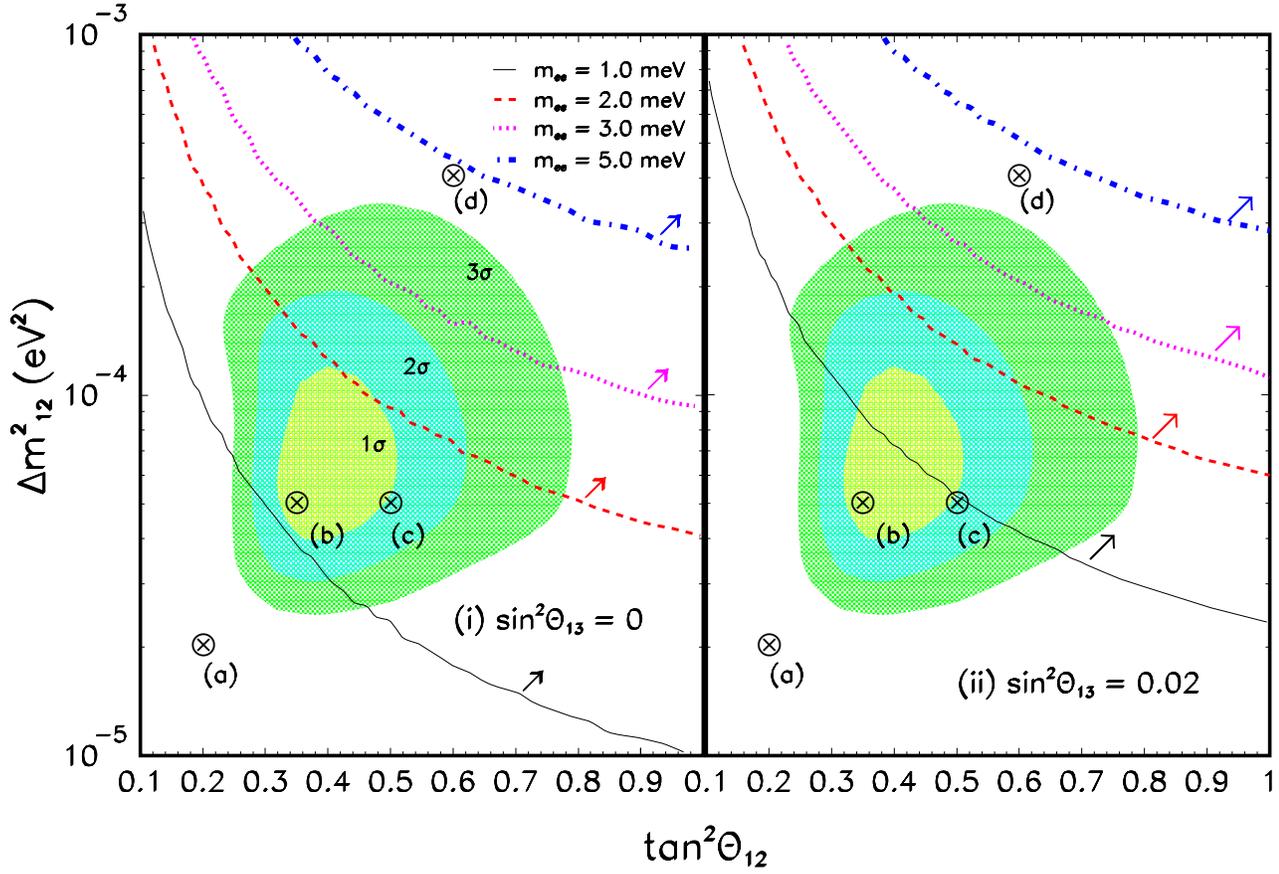}
\vglue -2.0cm
\caption{
Region in the ($\Delta m^2_{12}$, $\tan^2 \theta_{12}$) plane where
$\alpha_1$ can be constrained to a non-zero value, indicated by an
arrow, for given central values of $m_{ee}=1,2,3$ meV, taking into
account 30\% uncertainty in the determination of $m_{ee}$ and 10 \% 
uncertainty in $\Delta m^2_{12}$ as well as in $\tan^2 \theta_{12}$. 
The set of solar parameters used in Fig.~\ref{fig:m_ee-alpha1-m_0_A}(a)-(d)  
are indicated by crosses. 
The allowed region for the LMA MSW solution at 1, 2 and 3 $\sigma$ 
are shown by the shaded area (adopted from Ref.~\cite{Holanda}).
}
\label{fig:nova}
\end{figure}

\begin{figure}
\vglue 1.cm
\includegraphics[height=17.cm,width=16cm]{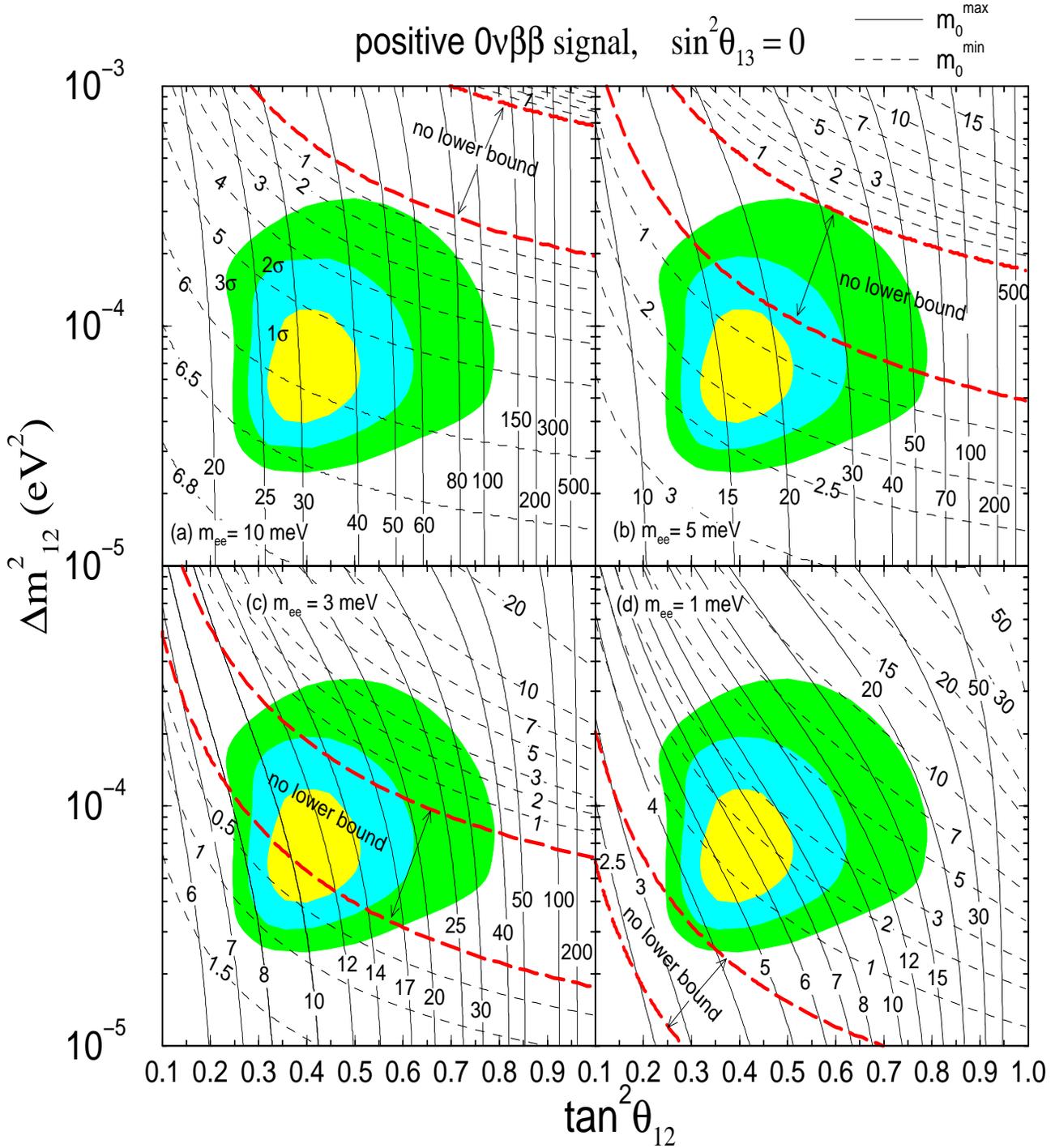}
\vglue 0.5cm
\caption{
Iso-contour plots of upper ($m_0^{\text{max}}$, solid curves) 
and lower ($m_0^{\text{min}}$, dashed curves) bounds of 
$m_0$ in units of meV in the $\tan^2\theta_{12}-\Delta m^2_{12}$ 
plane for the case where a positive signal of $0\nu\beta\beta$ 
is observed with central values, $m_{ee}$ = 10 meV (a), 5 meV (b), 
3 meV (c) and 1 meV (d).
We assume 30\% uncertainty in the determination of $m_{ee}$.  
We fix the other mixing parameters as 
$\Delta m^2_{23} = 3 \times 10^{-3}$ eV$^2$
and  $\sin^2 \theta_{13}=0$.
The allowed region for the LMA MSW solution is the same as in 
Fig.~\ref{fig:nova}.
}
\label{fig:m_0sol_norm_posi0}
\end{figure}

\begin{figure}
\vglue 1.cm
\includegraphics[height=17.cm,width=16cm]{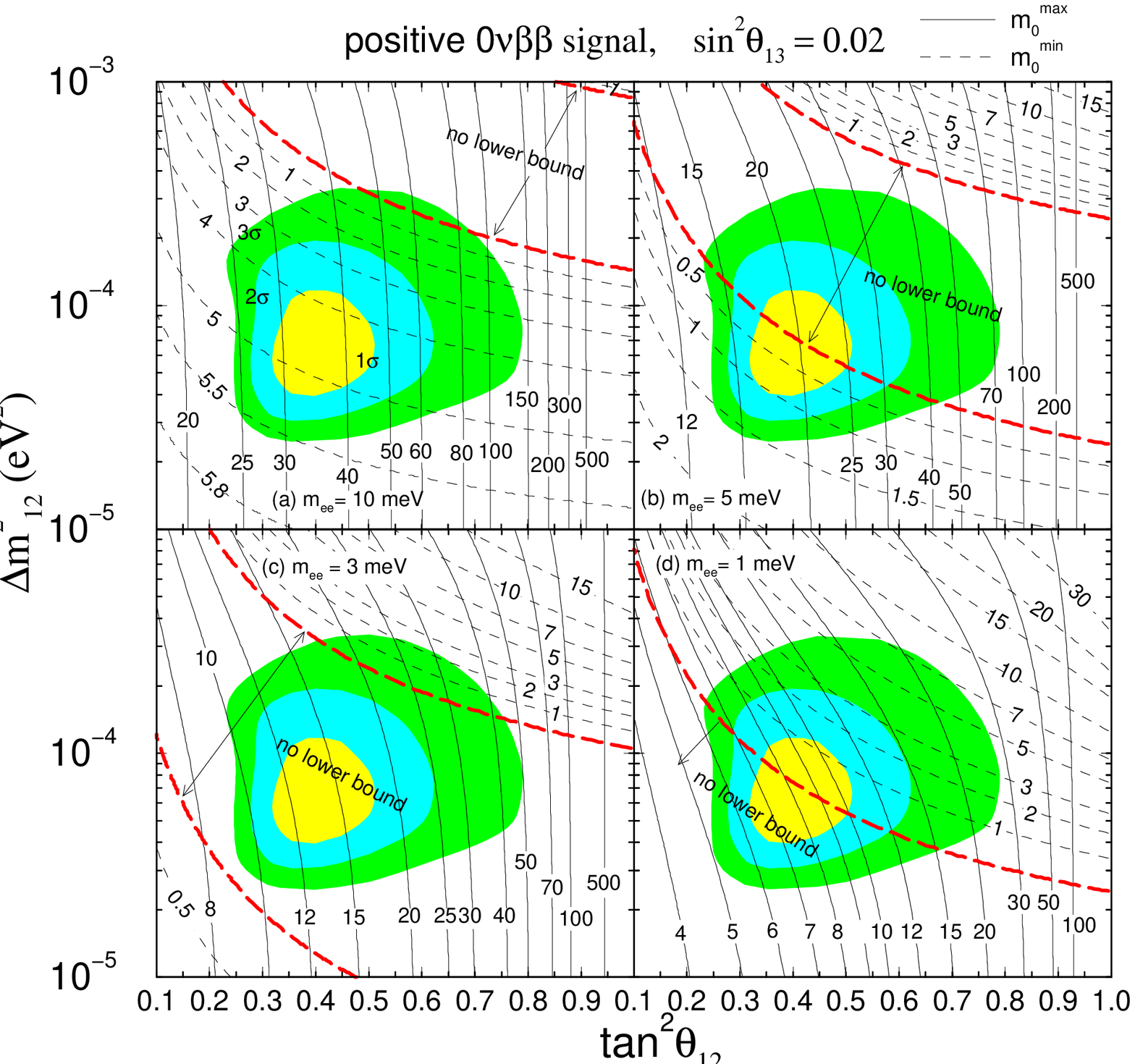}
\vglue 0.5cm
\caption{Same as Fig.~\ref{fig:m_0sol_norm_posi0} but for 
$\sin^2 \theta_{13}=0.02$.
}
\label{fig:m_0sol_norm_posi2}
\end{figure}

\begin{figure}
\vglue 1.cm
\includegraphics[height=17.cm,width=16cm]{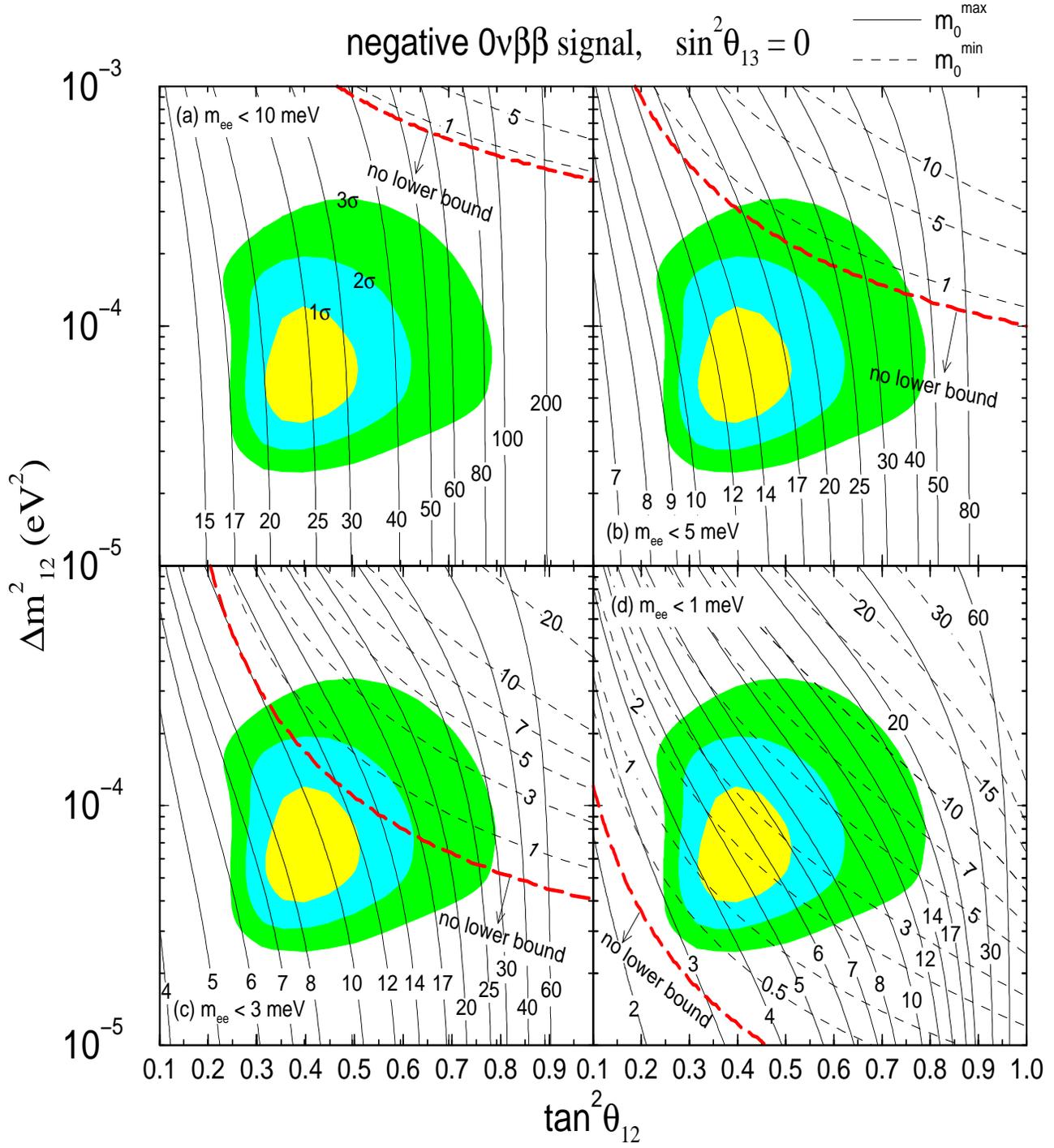}
\vglue 0.5cm
\caption{
Same as Fig.~\ref{fig:m_0sol_norm_posi0} but for 
the case where no positive signal of $0\nu\beta\beta$ 
is observed down to $m_{ee}$ = 10 meV (a), 5 meV (b), 
3 meV (c) and 1 meV (d). 
}
\label{fig:m_0sol_norm_nega0}
\end{figure}

\begin{figure}
\vglue 1.cm
\includegraphics[height=17.cm,width=16cm]{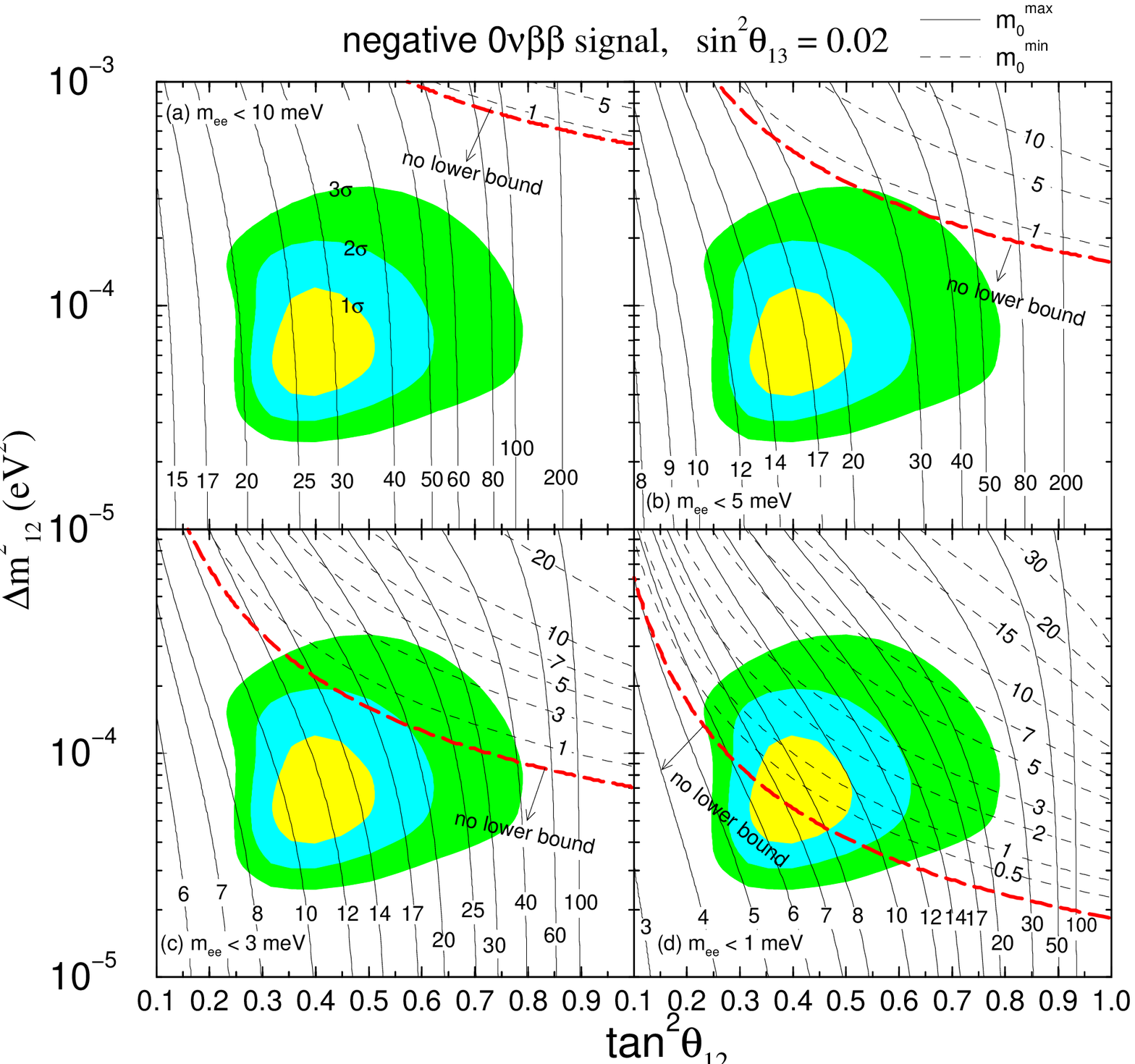}
\vglue 0.5cm
\caption{
Same as Fig.~\ref{fig:m_0sol_norm_nega0} but for 
$\sin^2 \theta_{13}=0.02$.
}
\label{fig:m_0sol_norm_nega2}
\end{figure}

 We finally conclude that it is possible to constrain the CP violating 
 phase $\alpha_1$ to values around $\pi/2$ if: 
 (1) $m_0$ is large enough to be detected by KATRIN or by astrophysical 
 observations and future $0\nu \beta \beta$ decay experiments observe 
 $m_{ee}$ close to its minimum value or (2) future $0\nu \beta \beta$ 
 decay experiments can achieve a sensitivity on $m_{ee} \lsim 5 $ meV,
 depending on the values of the solar parameters and on the
 uncertainty on $m_{ee}$ (Fig.~\ref{fig:nova}), independently of whether
 a positive or a negative signal is observed.
 Unfortunately, nothing can be known about $\alpha_3$.  
 
 Despite of the fact that the parameter region inspected in this work
 differs from the one considered in Ref.~\cite{barger}, 
 we have found, in agreement with this reference, that evidence for CP 
 violation cannot be observed by future $0 \nu \beta \beta$  decay 
 experiments, since the possible values of $\alpha_1$, 
 whenever it can be constrained to some non-zero value,
 always include  $\pi/2$, 
 unless $m_0$ can be independently determined by some 
 other experiment.

%%%%%%%%%%%%%% Thanks
\acknowledgments 
We thank P.~C.~de Holanda for useful correspondence. 
This work was supported by Funda\c{c}\~ao de Amparo
\`a Pesquisa do Estado de S\~ao Paulo (FAPESP) and by Conselho
Nacional de  Ci\^encia e Tecnologia (CNPq).

%%%%%%%%%%%%%%%%%%%%%%%%%%%%%%%%%%%%%%%%%%%%%%%%%%%%%%%%%%%%%%%%%%%%%%%
%%%%%%%%%%%%%% References 
%%%%%%%%%%%%%%%%%%%%%%%%%%%%%%%%%%%%%%%%%%%%%%%%%%%%%%%%%%%%%%%%%%%%%%%

%%%%%%%%%%%%%%%%%%%%%%%%%%%%%%%%%%%%%%%%%%%%%%%%%%%%%%%%%%%%%%%%%%%%%%%
%%%%%%%%%%%%%%%%%%%%%%%%%%%%%%%%%%%%%%%%%%%%%%%%%%%%%%%%%%%%%%%%%%%%%%%

\newpage

%%%%%%%%%%%%%%%%%%%%%%%%%%%%%%%%%%%%%%%%%%%%%%%%%%%%%%%%%%%%%%%%%%%%%%%
%%%%%%%%%%%%%% Figures
%%%%%%%%%%%%%%%%%%%%%%%%%%%%%%%%%%%%%%%%%%%%%%%%%%%%%%%%%%%%%%%%%%%%%%%

%%%%%%%%%%%%%%%%%%%%%%%%%%%%%%

%%%%%%%%%%%%%%%%%%%%%%%%%%%%%%

% \end{document}
\newpage

%%%%%%%%%%%%%%%%%%%%%%%%%%%%%%

\newpage
%%%%%%%%%%%%%%%%%%%%%%%%%%%%%%

\newpage
%%%%%%%%%%%%%%%%%%%%%%%%%%%%%%

\newpage
%%%%%%%%%%%%%%%%%%%%%%%%%%%%%%

%%%%%%%%%%%%%%%%%%%%%%%%%%%%%%
\newpage

%%%%%%%%%%%%%%%%%%%%%%%%%%%%%%
\newpage

%%%%%%%%%%%%%%%%%%%%%%%%%%%%%%
\newpage

%%%%%%%%%%%%%%%%%%%%%%%%%%%%%%
\newpage

%%%%%%%%%%%%%%%%%%%%%%%%%%%%%%
\newpage


\begin{thebibliography}{100}

\bibitem{solarnu} 
Q.~R.~Ahmad {\em et al.}  (SNO Collaboration),
Phys.\ Rev.\ Lett.\ {\bf 87}, 071301 (2001);
%
S.\ Fukuda {\em et al.} (Super-Kamiokande Collaboration),
Phys. Rev. Lett. {\bf 86},  5651 (2001);
{\it ibid.}  {\bf 86}, 5656 (2001); arXiv:hep-ex/0205075;
%
K. Lande {\em et al.} (Homestake Collaboration), 
Astrophys\ .J.\ {\bf 496}, 505 (1998); 
%
J. Abdurashitov {\em et al.} (SAGE Collaboration), 
Phys.\ Rev.\ C {\bf 60}, 055801 (1999); 
% 
W.\ Hampel {\em et al.} (GALLEX Collaboration), 
Phys.\ Lett.\  B {\bf447}, 127 (1999); 
%
M. Altmann {\em et al.} (GNO Collaboration),
Phys. Lett. B {\bf 490}, 16 (2000). 

%%% atmospheric 
\bibitem{atmnu} 

Y. Fukuda {\em et al.} (Super-Kamiokande Collaboration), 
Phys. Rev. Lett. {\bf 81}, 1562 (1998);
%
H. S. Hirata {\em et al.} (Kamiokande Collaboration), 
Phys. Lett. B {\bf 205}, 416 (1988); 
{\it ibid.}\ {\bf 280}, 146 (1992); 
Y. Fukuda {\em et al.}, {\it ibid.}\  {\bf 335}, 237 (1994); 
%
R. Becker-Szendy {\em et al.} (IMB Collaboration), 
Phys. Rev. D {\bf 46}, 3720 (1992); 
%
M. Ambrosio {\em et al.} (MACRO Collaboration), 
Phys. Lett. B {\bf 478}, 5 (2000); 
B. C. Barish,  Nucl. Phys. B (Proc. Suppl.) {\bf 91}, 141 (2001); 
%
W. W. M. Allison {\em et al.} (Soudan-2 Collaboration), 
Phys. Lett. B {\bf 391}, 491 (1997); 
Phys. Lett. B {\bf 449}, 137 (1999); 
W. A. Mann, Nucl. Phys. B (Proc. Suppl.) {\bf 91}, 134 (2001);
%

\bibitem{SNONC} 
Q.~R.~Ahmad {\em et al.}  (SNO Collaboration),
%Phys.\ Rev.\ Lett.\ {\bf 87}, 071301 (2001);
arXiv:nucl-ex/0204008, arXiv:nucl-ex/0204009.

%%%% K2K 
\bibitem{k2k}
S.~H.~Ahn {\em et al.}  (K2K Collaboration),
%``Detection of accelerator produced neutrinos at a distance of 250-km,''
Phys.\ Lett.\ B {\bf 511}, 178 (2001); 
%[arXiv:hep-ex/0103001].
%%CITATION = HEP-EX 0103001;%%
%
%\cite{Hill:2001gu}
% \bibitem{Hill:2001gu}
J.~E.~Hill  (K2K Collaboration),
%``Results from the K2K long-baseline neutrino oscillation experiment,''
in {\it Proc. of the APS/DPF/DPB Summer Study on the Future of Particle Physics 
(Snowmass 2001) } ed. by R.~Davidson and C.~Quigg, arXiv:hep-ex/0110034.
%%CITATION = HEP-EX 0110034;%%

%%%% CHOOZ

\bibitem{reactors} 
M. Apollonio {\em et al.} (CHOOZ Collaboration), 
Phys.\ Lett.\ B {\bf 420}, 397 (1998); 
{\it ibid.} {\bf 466}, 415 (1999);
%
%\cite{Boehm:2000vp}
%\bibitem{Boehm:2000vp}
F.~Boehm {\em et al.} (Palo Verde Collaboration), 
%``Results from the Palo Verde neutrino oscillation experiment,''
Phys.\ Rev.\ D {\bf 62}, 072002 (2000); 
%[arXiv:hep-ex/0003022].
%%CITATION = HEP-EX 0003022;%%
%F.~Boehm {\em et al.},
%``Final results from the Palo Verde neutrino oscillation experiment,''
Phys.\ Rev.\ D {\bf 64}, 112001 (2001).
%[arXiv:hep-ex/0107009].
%%CITATION = HEP-EX 0107009;%%


\bibitem{MNS} 
Z. Maki, M. Nakagawa, and S. Sakata,  
Prog. Theor.  Phys. {\bf 28}, 870 (1962). 

\bibitem{oscillation} 
H.~Minakata and H.~Nunokawa, JHEP {\bf 0110}, 001 (2001), 
V.~Barger {\em et al.}, Phys.\ Rev.\ D {\bf 65}, 053016 (2002);
T.~Kajita, H.~Minakata and H.~Nunokawa, Phys.\ Lett.\ B {\bf 528}, 245 (2002);
M.~Aoki {\em et al.}, arXiv:hep-ph/0112338;
P.~Huber, M.~Lindner and W. Winter, arXiv:hep-ph/0204352;
A.~Donini, D.~Meloni and P.~Migliozzi, arXiv:hep-ph/0206034;
V.~Barger, D.~Marfatia and K.~Whisnant, arXiv:hep-ph/0206038;
G.~Barenboim {\em et al.}, arXiv:hep-ph/0204208; arXiv:hep-ph/0206025.
  
\bibitem{kamland} 
V.~Barger, D.~Marfatia and B. P. Wood, Phys.\ Lett.\ B {\bf 498}, 53 (2001);
H.~Murayama and A.~Pierce, Phys.\ Rev.\ D {\bf 65}, 013012 (2002);
A.~de Gouvea and C.~Pena-Garay, Phys.\ Rev.\ D {\bf 65}, 0113011 (2001);
M.~C.~Gonzalez-Garcia and C.~Pena-Garay, Phys.\ Lett.\ B {\bf 527}, 199 (2002);
P. Aliani {\em et al.}, arXiv:hep-ph/0205061. 

\bibitem{Schechter} 
J.~Schechter and J.~W.~Valle,
%``Neutrino Masses In SU(2) X U(1) Theories,''
Phys.\ Rev.\ D {\bf 22}, 2227 (1980); 
%%CITATION = PHRVA,D22,2227;%%
%
%J.~Schechter and J.~W.~Valle,
%%``Neutrino Oscillation Thought Experiment,''
%Phys.\ Rev.\ D {\bf 23}, 1666 (1981);
%%CITATION = PHRVA,D23,1666;%%
%
{\it ibid.}  {\bf 23}, 1666 (1981);
%.~Schechter and J.~W.~Valle,
%``Majorana Neutrinos And Magnetic Fields,''
%Phys.\ Rev.\ D {\bf 24}, 1883 (1981)
%[Erratum-ibid.\ D {\bf 25}, 283 (1982)].
%%CITATION = PHRVA,D24,1883;%%
{\it ibid.}  {\bf 24}, 1883 (1981); 
[Erratum-ibid.\ D {\bf 25}, 283 (1982)].

\bibitem{Bilenky} 
S.~M.~Bilenky, J.~Hosek and S.~T.~Petcov,
%``On Oscillations Of Neutrinos With Dirac And Majorana Masses,''
Phys.\ Lett.\ B {\bf 94}, 495 (1980).
%%CITATION = PHLTA,B94,495;%%


\bibitem{Doi}
M.~Doi {\em et al.}, 
%M.~Doi, T.~Kotani, H.~Nishiura, K.~Okuda and E.~Takasugi,
%``CP Violation In Majorana Neutrinos,''
Phys.\ Lett.\ B {\bf 102}, 323 (1981).
%%CITATION = PHLTA,B102,323;%%

\bibitem{old} 
M. Doi, T. Kotani and E. Takasugi, 
Prog. Theor.  Phys. Suppl. {\bf 83}, 1 (1985); 
T. Tomoda, Rep. Prog. Phys. {\bf 54}, 53 (1991).
%
%

\bibitem{shechter} 
J.~Schechter and J.~W.~F.~Valle,  
Phys.\ Rev.\ D {\bf 25}, 2951 (1982). 



\bibitem{susy} 
M. Hirsch, H.~V.~Klapdor-Kleingrothaus and S.~G.~Kovalenko,
Phys.\ Rev.\ Lett. {\bf 75}, 17 (1995);
Phys.\ Rev.\ D {\bf 53}, 1329 (1996);
M.~Hirsch and J.~W.~F.~Valle, Nucl.\ Phys.\ B {\bf 557}, 60 (1999). 


\bibitem{0nuBB}

S.~T.~Petcov and A.~Yu.~Smirnov, Phys.\ Lett.\ B {\bf 322}, 109 (1994);
%
S.~M.~Bilenky, C.~Giunti, C.~W.~Kim and S.~T.~Petcov,
%``Short-baseline neutrino oscillations and neutrinoless double-beta decay in schemes with an inverted mass spectrum,''
Phys.\ Rev.\ D {\bf 54}, 4432 (1996)
%[arXiv:hep-ph/9604364].
%%CITATION = HEP-PH 9604364;%%
%
H.~Minakata and O.~Yasuda,
%``Constraining almost degenerate three-flavor neutrinos,''
Phys.\ Rev.\ D {\bf 56}, 1692 (1997); 
%
H.~Minakata and O.~Yasuda,
%``Dark matter neutrinos must come with degenerate masses,''
Nucl.\ Phys.\ B {\bf 523}, 597 (1998); 
%
T.~Fukuyama, K.~Matsuda and H.~Nishiura,
%``CP violation in neutrinoless double beta decay and neutrino  oscillation,''
Phys.\ Rev.\ D {\bf 57}, 5844 (1998); 
{\it ibid.}  {\bf 81}, 4279 (1998); 
%
F.~Vissani,
%``Signal of neutrinoless double beta decay, neutrino spectrum and  oscillation scenarios,''
JHEP {\bf 9906}, 022 (1999); 
%
%S.~M.~Bilenky, C.~Giunti, W.~Grimus, B.~Kayser and S.~T.~Petcov,
S.~M.~Bilenky {\em et al.}, 
%``Constraints from neutrino oscillation experiments on the effective  Majorana mass in neutrinoless double beta decay,''
Phys.\ Lett.\ B {\bf 465}, 193 (1999);
%
K.~Matsuda {\em et al.}, 
%``CP violations in lepton number violation processes and neutrino  oscillations,''
Phys.\ Rev.\ D {\bf 62}, 093001 (2000); 
%
H.~V.~Klapdor-Kleingrothaus, H.~P\"as and A.~Yu.~Smirnov,
%``Neutrino mass spectrum and neutrinoless double beta decay,''
Phys.\ Rev.\ D {\bf 63}, 073005 (2001);
%
W.~Rodejohann, Nucl.\ Phys.\ B {\bf 597}, 110 (2001); 
%
Y. Farzan, O. L. G. Peres and A. Yu. Smirnov, 
Nucl.\ Phys.\ B {\bf 612}, 59 (2001);
%
S. M. Bilenky, S. Pascoli and S. T. Petcov, 
Phys.\ Rev.\ D {\bf 64}, 053010 (2001);
%
H.~Minakata and H.~Sugiyama, Phys.\ Lett.\ B {\bf 526}, 335 (2002);
{\it ibid.} {\bf 532}, 275 (2002); 
%
.~Falcone and F.~Tramontano,
%``Neutrino oscillations and neutrinoless double beta decay,''
Phys.\ Rev.\ D {\bf 64}, 077302 (2001); 
%
Z.~z.~Xing,
%``Model-independent constraint on the neutrino mass spectrum from the neutrinoless double beta decay,''
Phys.\ Rev.\ D {\bf 65}, 077302 (2002);
%
H.~J.~He, D.~A.~Dicus and J.~N.~Ng,
%``Minimal schemes for large neutrino mixings with inverted hierarchy,''
Phys.\ Lett.\ B {\bf 536}, 83 (2002)
%[arXiv:hep-ph/0203237].
%
S.~Pascoli, S.~T.~Petcov and L.~Wolfenstein,
%``Searching for the CP-violation associated with Majorana neutrinos,''
Phys.\ Lett.\ B {\bf 524}, 319 (2002)
%[arXiv:hep-ph/0110287].
%%CITATION = HEP-PH 0110287;%%
%
W.~Rodejohann, arXiv:hep-ph/0203214; 
%
F.~Feruglio, A.~Strumia and F.~Vissani,  
arXiv:hep-ph/0201291, to appear in Nucl.\ Phys.\ B;
%
S. Pascoli and S. T. Petcov, 
arXiv:hep-ph/0205022;
%
S.~Pascoli and S.~T.~Petcov,
%``Majorana neutrinos, CP-violation, neutrinoless double-beta and  tritium-beta decays,''
arXiv:hep-ph/0111203.
%%CITATION = HEP-PH 0111203;%%


\bibitem{0nuBB-absolute}
For instance, the following articles also discussed and demonstrated 
the relation between $0\nu\beta\beta$ decay signal and the absolute 
neutrino mass scales and/or Majorana CP phases but in different ways 
from the ones we consider in this work:
%
V.~Barger and K. Whisnant, Phys.\ Lett.\ B {\bf 456}, 194 (1999);
H.~P\"as and T. J. Weiler, Phys.\ Rev.\ D {\bf 63}, 113015 (2001);
P.~Osland and G.~Vigdel, Phys.\ Lett.\ B {\bf 520}, 143 (2001);
V.~Barger {\em et al.}, Phys.\ Lett.\ B {\bf 532}, 15 (2002);
M.~Frigerio and A.~Yu.~Smirnov,
%``Structure of neutrino mass matrix and CP violation,''
arXiv:hep-ph/0202247. 
%%CITATION = HEP-PH 0202247;%%


\bibitem{upperbound} 
H.~V.~Klapdor-Kleingrothaus {\em et al.}
(Heidelberg--Moscow Collaboration),
%``Latest results from the Heidelberg-Moscow double-beta-decay experiment,''
Eur.\ Phys.\ J. A {\bf 12}, 147 (2001);
%\bibitem{Aalseth:2002rf}
see also C.~E.~Aalseth {\em et al.}  (16EX Collaboration),
%``The IGEX Ge-76 neutrinoless double-beta decay experiment: Prospects for  next generation experiments,''
arXiv:hep-ex/0202026.


\bibitem{klapdor01} 
H.~V.~Klapdor-Kleingrothaus {\em et al.}, 
Mod.\ Phys.\ Lett.\ A {\bf 16}, 2409 (2001). 

\bibitem{comments} 
C.~E.~Aalseth  {\em et al.}, arXiv:hep-ex/0202018;
H.~V.~Klapdor-Kleingrothaus, arXiv:hep-ph/0205288;
H.~L.~Harney, arXiv:hep-ph/0205293.

\bibitem{GENIUS} 
H.~V.~Klapdor-Kleingrothaus {\em et al.} (GENIUS Collaboration),
arXiv:hep-ph/9910205. 

\bibitem{CUORE} 
E.~Fiorini {\em et al.}, Phys.\ Rep.\ {\bf 307}, 309 (1998);
A.~Bettini, Nucl.\ Phys.\ Proc. Suppl. {\bf 100}, 332 (2001). 

\bibitem{EXO} 
M. Danilov {\em et al.}, Phys.\ Lett.\ B {\bf 480}, 12 (2000).

\bibitem{majorana} 
C. E. Aalseth {\em et al.} (MAJORANA Collaboration), 
arXiv:hep-ex/0201021.

\bibitem{NOON} 
H. Ejiri {\em et al.}, Phys.\ Rev.\ Lett. {\bf 85}, 2917 (2000). 

\bibitem{mainz} 
J.~Bonn {\em et al.}, (Mainz Collaboration), 
Nucl.\ Phys.\ Proc. Suppl. {\bf 91}, 273 (2001).

\bibitem{Katrin} 
A.~Osipowicz {\em et al.} (KATRIN Collaboration), 
arXiv:hep-ex/0109033.

\bibitem{sol-recent} 
G. L. Fogli {\em et al.}, Phys. Rev. D {\bf 64}, 093007 (2001);
A.~M.~Gago {\em et al.},  Phys. Rev. D {\bf 65}, 073012 (2002);
%\bibitem{sol-new} 
J.~N. Bahcall {\em et al.}, arXiv:hep-ph/0204314;
A.~Bandyopadhyay {\em et al.}, arXiv:hep-ph/0204286
%
V.~Barger {\em et al.}, arXiv:hep-ph/0204253;
%
P.~C.~de Holanda and A.~Yu.~Smirnov arXiv:hep-ph/0205241;
%
P.~Creminelli {\em et al.}, arXiv:hep-ph/0102234.

\bibitem{vogel} 
For a recent review on $0\nu \beta\beta$ decay, 
see S.~R.~Elliot and P.~Vogel, arXiv:hep-ph/0202264.


\bibitem{barger} 
V.~Barger {\em et al.}, arXiv:hep-ph/0205290.

\bibitem{Holanda} 
See P.~C.~de Holanda and A.~Yu.~Smirnov in Ref~\cite{sol-recent}. 

\bibitem{MAP} 
See http://map.gsfc.nasa.gov/

\bibitem{Planck} 
See http://astro.estec.esa.nl/SA-general/Projects/Planck/.

\bibitem{cosmology} 
W. Hu, D. J. Eisenstein and M. Tegmark,
Phys. Rev. Lett. {\bf 80}, 5255 (1998); 
%
M. Tegmark, M. Zaldarriaga and A. J. Hamilton, 
Nucl.\ Phys.\ Proc. Suppl. {\bf 91}, 38 (2001).

 \bibitem{supernova} 
 T. Totani, Phys. Rev. Lett. {\bf 80}, 2039 (1998);
 J. F. Beacom, R. N. Boyd and A. Mezzacappa, 
 Phys. Rev. Lett. {\bf 85}, 3568 (2000). 

\end{thebibliography}
\end{document}